\documentclass[a4paper,12pt]{article}
\usepackage{amsmath}
\usepackage{amssymb}
\usepackage{graphicx}
\usepackage[usenames]{color}
\usepackage{latexsym}
\usepackage{mathrsfs}
\def\al{\alpha}
\def\be{\beta}
\def\ga{\gamma} \def\Ga{\Gamma}
\def\ep{\epsilon}
\def\lam{\lambda}

\def\th{\theta}
\def\sig{\sigma}
\def\Th{\Theta}
\def\alp{{\alpha'}}

\def\sigp{{\sigma'}}
\def\parn              {  \par\noindent }

\def\calA{{\cal A}}   
  
 \def\calH{{\cal H}} 
 \def\calK{{\cal K}} \def\calL{{\cal L}}
 \def\calN{{\cal N}} \def\calO{{\cal O}}
\def\calP{{\cal P}} \def\calQ{{\cal Q}} 
 \def\calT{{\cal T}} 
  \def\calX{{\cal X}}
  
\def\matrixii#1#2#3#4            {  \left(\begin{array}{cc}#1&#2\\#3&#4
                                       \end{array}\right) }
\def\vecii#1#2      {  \left(\begin{array}{c}#1\\#2\end{array}\right)  }
\def\del        {  \partial }
\def\half       {  {1\over 2}  }

\def\abs#1      {  \vert #1 \vert  }
\def\ie         {  {{\it i.e.}\  }    }

\def\comma          {\, ,}
\def\period         {\, .}
\def\lsim      {\lower .65ex \hbox{\ $\stackrel{<}{\sim}$\ } }
\def\gsim      {\lower .65ex \hbox{\ $\stackrel{>}{\sim}$\ } }

\def\ket#1{{| #1 \rangle } }
\def\com#1#2{{ \left[#1, #2\right] } } 
\def\acom#1#2{{ \left\{ #1,#2\right\} } }
\def\nn               {  \nonumber  }
\newcommand{\nullify}[1]{}
\def\atil{\tilde{a}}

\def\btil{\tilde{b}}

\def\Qtil{\widetilde{Q}}

\def\Stil{\widetilde{S}}
\def\util{\tilde{u}}

\def\altil{\tilde{\alpha}}

\def\deltil{\tilde{\del}}

\def\ttil{\tilde{t}}
\def\omegatil{\tilde{\omega}}

\def\lamtil{\tilde{\lambda}}

\def\sigtil{\tilde{\sigma}}

\def\Pitil{\widetilde{\Pi}}

\def\Pitilhat{\hat{\widetilde{\Pi}}}
\def\adot{{\dot{a}}}

\def\muhat{\hat{\mu}}

\def\pphat{\hat{p}^+}

\def\Pihat{\hat{\Pi}}
\def\Shat{\hat{S}}
\def\Shatonep{\hat{S}^{1 (+)}}
\def\Shatonem{\hat{S}^{1 (-)}}
\def\calThat{\hat{\cal{T}}}

\def\gabar{\bar{\gamma}}

\def\pp{{p^+}}

\def\sp{\sigma_+}
\def\sm{\sigma_-}

\def\pplusup{p^+}
\def\pminusup{p^-}

\def\Xp{X^+}
\def\Xm{X^-}

\def\lamp{\lambda^+}
\def\lamm{\lambda^-} 
\def\lamtilp{\tilde{\lambda}^+}
\def\lamtilm{\tilde{\lambda}^-}
\def\gabarm{\bar{\gamma}^-}

\def\piplus#1{\pi^{+#1}}

\def\Omp{\Omega^+}

\def\calXpl{\calX^+_L}
\def\calXpr{\calX^+_R}

\def\inv{^{-1}}

\def\qinv{q^{-1}}

\def\Ncheck{\check{N}}

\def\Pcom#1#2{\left\{#1,#2\right\}_P}
\def\Dcom#1#2{\left\{#1, #2\right\}_D}

\def\deltassp{\delta(\sig -\sigp)} 
\def\deltapssp{\delta'(\sig-\sigp)}
\def\deltasspep{\delta_\ep(\sig-\sigp)} 
\def\deltapsspep{\delta'_\ep(\sig-\sigp)} 

\def\toplus{^{(+)}} 
\def\tominus{^{(-)}}
\def\varth{\vartheta}

\def\sqtwo{\sqrt{2}}
 
\def\ls{\ell_s}

\def\lcvac{\ket{0}_{lc}}

\def\circbox{\hbox{$\scriptscriptstyle\circ$}}
\def\timesbox{\hbox{$\scriptscriptstyle\times$}}
\def\anc{{{\lower 1ex \circbox} \atop {\raise 1.5ex \circbox}}}
\def\ant{ {{\lower 1ex  \timesbox} \atop {\raise 1.5ex  \timesbox}}}
\def\ap{{\raise 0.2ex \hbox{$\scriptstyle \odot$}}}
\def\Ker{{\rm Ker}\,}
\def\papertitlepage{\baselineskip 3.5ex \thispagestyle{empty}}
\def\Title#1{\baselineskip 1cm \vspace{1.5cm}\begin{center}
 {\Large\bf #1} \end{center} 
\vspace{0.5cm}}
\def\Authors#1{\begin{center} {\it #1} \end{center}}
\def\Abstract{\vspace{1.0cm}\begin{center} {\large\bf Abstract} 
           \end{center} \par\bigskip}
\def\Komabanumber#1#2{\hfill \begin{minipage}{4.2cm} UT-Komaba #1
              \parn #2 
              \end{minipage}}
\renewcommand{\thefootnote}{\fnsymbol{footnote}}
\renewenvironment{thebibliography}{\pagebreak[3]\par\vspace{0.6em}
\begin{flushleft}{\large \bf References}\end{flushleft}
\vspace{-1.0em}

\begin{enumerate}\if@twocolumn\baselineskip=0.6em\itemsep -0.2em
\else\itemsep -0.2em\fi\labelsep 0.1em}{\end{enumerate} }

\begin{document}
\papertitlepage
\vspace*{0cm}
\Komabanumber{08-2}{January, 2008}
\Title{Superstring in the plane-wave background  \\
with RR  flux 
as a conformal field theory} 
\Authors{{\sc Yoichi Kazama\footnote[2]{kazama@hep1.c.u-tokyo.ac.jp} 
 and Naoto Yokoi\footnote[3]{nyokoi@hep1.c.u-tokyo.ac.jp}
\\ }
\vskip 3ex
 Institute of Physics, University of Tokyo, \\
 Komaba, Meguro-ku, Tokyo 153-8902 Japan \\
  }
\baselineskip .7cm
\numberwithin{equation}{section}
\numberwithin{figure}{section}
\numberwithin{table}{section}
\parskip=0.9ex
\Abstract
We  study  the type IIB superstring in the plane-wave 
background  with Ramond-Ramond flux and formulate it as an exact conformal field theory in operator formalism. One of the characteristic features of the superstring in a consistent background with  RR flux, such as the $AdS_5 \times S^5$ and  its plane-wave limit,  is that the left- and the right-moving degrees  of freedom on the worldsheet are inherently coupled.
 In the plane-wave  case, it is manifested in the well-known fact that the Green-Schwarz formulation  of the theory reduces to that of free massive bosons and fermions  in the light-cone gauge. This raises the obvious question as to  how this feature is reconciled with the underlying conformal symmetry of the string theory. By adopting  the semi-light-cone conformal gauge,  we will show that,  despite  the existence of  such non-linear left-right couplings, one can construct  two independent sets of  quantum Virasoro operators  in terms of fields obeying  the free-field commutation relations. Furthermore, we  demonstrate that the BRST cohomology analysis reproduces   the physical spectrum obtained  in the light-cone gauge. 
\baselineskip 3.5ex
\section{Introduction and summary}  
\renewcommand{\thefootnote}{\arabic{footnote}}
Undoubtedly, the AdS/CFT correspondence\cite{Maldacena:1997re,Gubser:1998bc,
Witten:1998qj} is 
one of the most profound  structures in string theory. In the past 10 years, an impressive
 collection of evidences 
 have been accumulated in favor of  this remarkable conjecture. On the CFT side, fairly 
 detailed analyses have  recently become possible, in particular  for the $N=4$ super-Yang-Mills
 theory, with the powerful assumption of integrability\cite{Minahan:2002ve, Beisert:2003tq, 
Beisert:2003yb, Beisert:2004ry} as well as with the state of the art 
perturbative  techniques\cite{Bern:2005iz, Bern:2006ew, Bern:2007ct}. 
On the other hand, the corresponding developments on the AdS side have 
 been available  mostly in the classical or semi-classical 
regimes\cite{Gubser:2002tv, Mandal:2002fs, Bena:2003wd, Tseytlin:2004xa}. Understanding of the 
stringy aspects has been comparatively slow due to the difficulty of solving the string theory 
 in the relevant curved backgrounds  with large Ramond-Ramond(RR) flux. In any case, the fundamental mechanism of this correspondence is yet to be unravelled. 

There are  strong   reasons   to believe that the presence of the RR flux must play   a key   role 
 in this strong-weak duality. Most directly, the basic relation 
$4\pi g_s N = g_{YM}^2 N = R^4/\alp^2$ 
expresses  the balance between the curvature and the RR flux,  which keeps the 
$AdS_5 \times S^5$ type curved geometry from gravitationally collapsing. Unfortunately, 
however, the treatment of the RR flux (together with the curved background) is precisely 
 the major obstacle which has been hampering the progress of the string side of the 
 AdS/CFT correspondence. 

One notable exception is the superstring in the 
plane-wave limit of $AdS_5 \times S^5$\cite{Penrose:1976, Blau:2001ne, Blau:2002dy}. 
As was first shown in \cite{Metsaev:2001bj}, by adopting the light-cone gauge in the 
Green-Schwarz formalism\cite{Green:1983wt, Green:1983sg, Metsaev:1998it}, 
the worldsheet theory in this case becomes simply a collection of free massive bosons 
 and fermions. This fact was exploited  in \cite{Berenstein:2002jq} to initiate a detailed
  comparison of the spectrum of the energy of the  string and that 
of the anomalous dimensions 
 of the corresponding gauge-invariant operators in super-Yang-Mills theory. 
Subsequent surge of researches on this so-called BMN limit advanced our understanding 
 of the nature of the AdS/CFT correspondence enormously (see \cite{Plefka:2003nb, Sadri:2003pr} for reviews). 

However,  a number of important aspects of the string theory
 in this background are still to be clarified. For one thing, 
the analysis  of the interactions of the string 
is not  straightforward. Although the three point 
vertex has been  computed using the light-cone string field theory\cite{Spradlin:2002ar, Spradlin:2002rv, Constable:2002hw, Dobashi:2002ar, Gomis:2002wi, Pankiewicz:2002tg, DiVecchia:2003yp, Pankiewicz:2003ap, Dobashi:2004nm, Spradlin:2003xc},   higher point functions have not yet
 been constructed. Another important  aspect which needs to be better understood 
is the modular property. Because the modular $S$-transformation affects  the light-cone gauge condition itself, the partition function is not modular invariant 
 but only modular ``covariant" (in a certain sense)\cite{Bergman:2002hv, Takayanagi:2002pi}. Including these questions, one should understand the string theory in this background 
more fully as a precious  prototypical model, in particular for understanding the role 
 of the large RR flux in the AdS/CFT correspondence. 

One major factor which has been preventing  further understanding of the theory 
 is the lack of  conformal invariance in the light-cone gauge formulation. 
As a string theory, it should be possible to formulate it  as an exact conformal 
 field theory (CFT)
 and make use of its tight structures and poweful techniques\footnote{For the CFT formulation of a superstring in a class of  $AdS_3$  backgrounds with RR flux, there have been a large number of investigations. See for example 
 \cite{Pesando:1998wm, Rahmfeld:1998zn, Park:1998un, Berkovits:1999im,Kutasov:1999xu}. }. Once it is achieved 
in a tractable manner, we should be able to construct the interaction vertices more 
easily and discuss the modular invariance in a  proper setting. Also, it should serve as  
a basis for constructing an operator version of  the covariant pure spinor formalism  in this  background\cite{Berkovits:2000fe, Berkovits:2002zv, Berkovits:2002vn}. 

One would expect, however,  that such a CFT formulation 
 is not so straightforward. This is because the theory at hand has quite an 
unusual feature.  Namely, just as in 
 the  $AdS_5\times S^5$ background, 
the left- and the right-moving degrees of freedom on the worldsheet are 
coupled  in the plane wave background right from the beginning. 
In the light-cone gauge, this 
is manifested as the massive nature of the bosons and the fermions. On the other hand, 
 a CFT  description means, by definition, that the ``left" and the ``right"
sectors must ``decouple" in the sense that they form  representations of 
 two mutually commuting Virasoro algebras. It is extremely interesting to see how these two 
 features  are reconciled. Another obvious difficulty is that in the conformal gauge
  the action is no longer quadratic and even the classical analysis, let alone the 
 quantum extension,  would become  quite   non-trivial. 

In this paper,  we shall show that,  despite such  anticipated  obstacles, 
it is  possible to achieve  an exact  CFT description,  not only classically but 
 quantum-mechanically as well,  in an operator formalism. 
In the case with the NSNS flux, a similar CFT formulation 
 was recently achieved in the RNS formalism using the canonical quantization method\cite{Chizaki:2007sc, Chizaki:2006pq}.  As we shall  see, with the RR flux the left-right coupling is much more 
inherent  and such a canonical method encounters  a severe difficulty. We will  overcome 
this difficulty  by the phase-space formulation, which does not require the knowledge of the solutions of the  equations of motion. 

Let us now give a summary of our results, which at the same 
 time serves  to indicate the organization of the paper. 

We begin the  analysis at the classical level in Sec. 2. We take our 
  basic Lagrangian to be  the one  constructed by Metsaev\cite{Metsaev:2001bj} in the Green-Schwarz formalism in the semi-light-cone(SLC) conformal 
gauge\cite{Carlip:1986cy, Carlip:1986cz, Kallosh:1987vq}. 
As briefly reviewed in Sec.2.1, it is composed  of the string coordinates $X^\mu
=(X^+,. X^-, X^I), ( I=1\sim 8)$ and the two sets of 16-component 
 Majorana spinors $\th^A_\al, (A=1,2)$ and contains  quartic interactions
expressing the coupling to the curved space and to the RR flux through a ``mass" 
parameter $\mu$.

 For a pedagogical reason, we will first describe, in Sec.2.2, what happens if one tries to treat this system by the canonical method. Despite non-linearity, the equations of 
 motion can be solved and the general solutions for all the basic fields are obtained. 
However, at this stage one already encounters a sign of difficulty. Except for 
 $X^+$, which is a free field, all the other fields are expanded  in terms 
 of the basis functions, to be called  $u_n$ and $\util_n$, which not only 
 depend on the modes of $X^+$ but also on the worldsheet light-cone coordinates $\sig_\pm=t\pm \sig$ {\it inseparably} due to the presence of $\mu$. 

To see how this is compatible with conformal 
 invariance, we compute the energy momentum tensors $\calT_\pm(=T_{\pm\pm})$ 
and substitute the solutions of the equations of motion. We then find that indeed 
$\calT_\pm$ become functions of $\sig_\pm$ respectively but in a peculiar manner.  All the 
dependence on $X^I, X^-$ and $\theta^A_\al$ 
 disappear and $\calT_\pm$ collapse to  exceedingly  simple expressions  involving 
$X^+$ and unknown ``holomorphic functions" $f_\pm(\sig_\pm)$, which appear 
in the solution of  $X^-$. This situation is not smoothly connected to the flat case with 
 $\mu=0$. 
These functions are to be determined by the requirement 
 that the basic fields and their conjugate momenta must satisfy the canononical 
 Poisson bracket relations. 
The problem is that, although the Poisson brackets can be 
 defined in the usual manner, it is practically impossible to analyze what $f_\pm$ 
 should be, due to the highly complicated completeness relations for  the operator-valued basis functions  $u_n$ and $\util_n$. In this sense the canonical analysis ``fails" even at
 the classical level. 

To overcome this difficulty, we turn, in Sec.2.3, to  the phase-space formulation. Although the method itself is completely standard,  a crucial observation 
 is that for theories, such as a string theory,  where the Hamiltonian is a member 
of a large symmetry algebra, the dynamics can be encoded in the ``kinematics", 
namely its representation  theory. This allows us to focus on the Virasoro algebra structure {\it at one time slice}, say  at $t=0$,  without recourse to the knowledge of the equations of motion. In spite of 
 the presence of left-right couplings, one can indeed verify that $\calT_\pm$ satisfy   two mutually commuting 
sets  of Virasoro algebras  at the classical level. 

We then turn, in Sec.3, to the quantum analysis. As we employ the phase-space formulation, 
the quantization of the basic fields, described in Sec.3.1,  is  straightforward. 
A great advantage of our formulation is that although these quantized fields time-develop
non-trivially according to the full non-quadratic Hamiltonian, they obey the free-field commutation relations. What is non-trivial, however, is to find the appropriate normal-ordering 
 prescription for the quantum Virasoro operators. In Sec.3.1.1, we define what we will call 
 the phase-space normal-ordering and compute the commutators of the Virasoro operators. 
We find that,  with an addition of a suitable quantum correction, they form two independent
 sets of Virasoro algebas  with central charge 26, as desired. We also examine, in Sec.3.1.2, 
another scheme, to be called massless normal-ordering, which appears more natural 
 for $\mu=0$. Extending it to the $\mu\ne 0$ case, we find that the Virasoro commutators 
 produce  operator anomalies proportional to $\mu^2$, which cannot be removed by 
quantum corrections. Some details of these computations are displayed  in Appendix A. 

Having found   the quantum Virasoro operators, it is straightforward to construct the 
 BRST operator  and study its cohomology. This is performed in Sec.4.1. 
Just as in the usual free string theory, the 
physical states will be identified as those in the transverse Hilbert space $\calH_T$, {\it i.e.}  without 
the non-zero modes of $X^\pm$, their conjugates and the ghosts, satisfying 
the Hamiltonian and the momentum  constraints $H=P=0$. These constraints can be easily 
 diagonalized in terms of ``massive"  oscillators constructed out of the phase space 
fields  in Sec.4.2. The re-normal-ordering constants produced in this process cancel between 
the  bosons and the fermions and we precisely reproduce the spectrum
 obtained in the light-cone  gauge together with the level-matching condition. 

The final section, Sec.5, will be devoted to a  discussion of some issues to be further clarified 
and  future perspectives. 
\section{Classical analysis}
\subsection{The basic Lagrangian  in the semi-light-cone gauge}
The Lagrangian of the type IIB Green-Schwarz superstring in the plane-wave 
 background with RR flux was constructed  in \cite{Metsaev:2001bj, Metsaev:2002re}. 
The basic fields are the string coordinates $X^\mu = (X^+, X^-, X^I)$ (where $X^\pm \equiv {1\over \sqrt{2}} (X^9\pm X^0)$ and $I=1\sim 8$)
 and the two sets of 16-component Majorana spinors $\th^A_\al  (A=1,2)$ of the same 10 dimensional chirality. We will  essentially follow the 
convention  of \cite{Metsaev:2001bj} with slight  changes\footnote{In \cite{Metsaev:2001bj} Metsaev flipped  
 the sign of $X^\mu$ just before writing down 
 the form of the Lagrangian in the semi-light-cone gauge.
 We do not make this change and adopt his original convention.}. The $32 \times 32$ $\Ga^\mu$-matrices 
are decomposed into  $16\times 16$ $\ga^\mu$ and $\gabar^\mu$ matrices as
$\Ga^\mu =\matrixii{0}{\ga^\mu}{\gabar^\mu}{0} $, where $(\ga^\mu)^{\al\be}
=(1, \ga^I, \ga^9)^{\al\be}$ and $(\gabar^\mu)_{\al\be} = (-1, \ga^I, \ga^9)_{\al\be}$. 
A matrix $\Ga$ is defined as $\Ga^\al{}_\be \equiv (\ga^1\gabar^2\ga^3\gabar^4)^\al{}_\be$ and it enjoys the property $\Ga^2=1$. 
The worldsheet coordinates are denoted  by $\xi^i=(t,\sig), i=0,1$
 and we define $\sig_\pm  \equiv t \pm \sig$. The conventions  for the flat worldsheet metric $\eta_{ij}$ and the $\ep^{ij}$ tensor  are  $\eta_{ij} = (-1, +1)$ and $\ep^{01} =1$. 

After fixing the 
 $\kappa$ symmetry by imposing the semi-light-cone(SLC) gauge 
condition\cite{Carlip:1986cy, Carlip:1986cz, Kallosh:1987vq} 
\begin{align}
\gabar^+\theta^A&= 0 \comma \label{slccond}
\end{align}
the Lagrangian density  is given by 
\begin{align}
\calL &= \calL_{kin} + \calL_{WZ} \comma \\
\calL_{kin} &= -{T\over 2}\sqrt{-g} g^{ij} 
\left( 2\del_i \Xp \del_j \Xm -\mu^2 X_I^2 \del_i \Xp \del_j \Xp 
+ \del_i X^I \del_j X^I \right) \nn\\
& +iT \sqrt{-g} g^{ij} \left[\del_i  \Xp ( \th^1 \gabarm \del_j \th^1 
+\th^2 \gabarm \del_j \th^2)+2\mu \del_i \Xp \del_j \Xp  \th^1\gabarm
\Ga  \th^2 \right]\comma  \label{Lkin1} \\
\calL_{WZ} &= -iT\ep^{ij} \del_i \Xp( \th^1 \gabarm \del_j \th^1 
-\th^2 \gabarm \del_j \th^2)\period  \label{LWZ1}
 \end{align}
$\calL_{kin}$ is the kinetic part and  $\calL_{WZ}$ is the Wess-Zumino part. 
The quartic terms in $\calL_{kin}$  proportional to $\mu^2$ and $\mu$  describe, respectively,   the coupling to the curved geometry  and to the RR flux. 
We will keep the  parameter $\mu$ carrying  the dimension of mass throughout. 
We also keep the string scale explicitly. It is expressed through either the slope 
 parameter $\alp$, the string tension
 $T=1/2\pi \alp$,  or the string length $\ls = \sqrt{\alp/2}$. 

Due to the SLC gauge condition, we can reduce the fermions to 8-component 
$SO(8)$ spinors and  simplify the Lagrangian. In the basis where $\gabar^9$
 is diagonal,  the spinor is decomposed as $\theta^A_\al= (\th^A_a, \th^A_\adot)$, 
and the SLC condition eliminates $SO(8)$-anti-chiral components $\th^A_\adot$. 
Furthermore, since the matrix $\Ga$  commutes with $\gabar^9$, it can also be 
restricted to the $SO(8)$-chiral sector, retaining its basic properties $\Ga_{ab}\Ga_{bc}=\delta_{ac}$ and $\Ga_{ab} = \Ga_{ba}$. Hence, we may make a redefinition
 $\Ga\th^2 \rightarrow \th^2$ and eliminate $\Ga$ altogether. The resultant 
 Lagrangian reads
\begin{align}
\calL_{kin} &= -{T\over 2}\sqrt{-g} g^{ij} 
\left( 2\del_i \Xp \del_j \Xm -\mu^2 X_I^2 \del_i \Xp \del_j \Xp 
+ \del_i X^I \del_j X^I \right) \nn\\
& +i\sqrt{2} T \sqrt{-g} g^{ij} \left[\del_i  \Xp ( \th^1 \del_j \th^1 
+\th^2 \del_j \th^2)+2\mu \del_i \Xp \del_j \Xp  \th^1
 \th^2 \right]\comma  \label{Lkin2} \\
\calL_{WZ} &= -i\sqrt{2} T\ep^{ij} \del_i \Xp( \th^1  \del_j \th^1 
-\th^2 \del_j \th^2)\comma   \label{LWZ2}
\end{align}
where $\th^1\del_j \th^1 \equiv  \th^1_a \del_j \th^1_a$, $\th^1\th^2 \equiv  \th^1_a \th^2_a$, etc. This is the form to be used in the subsequent analysis. 
\subsection{Canonical analysis }
The usual method for quantizing a field theory is to first find  the complete set
 of solutions of the equations of motion and then set up the commutation 
relations among  the time-independent coefficients in those solutions in such a way to  realize the canonical equal-time commutation  relations for  the basic fields. In the case 
 of  string theory, one must , in addition, identify 
 the Virasoro constraints and express
 them  in terms of such  quantized fields. 
In this subsection, 
we will describe what happens if one follows this canonical path. 
We will see that one encouters  some unusual features for the system at hand. 
\subsubsection{Equations of motion and their solutions }
Let us begin with the analysis of  the equations of motion,  with $g_{ij}$ set equal to
$\eta_{ij}$. The simplest is 
 the equation for the field $X^+$. By varying $\calL$ with respect to $X^-$ one obtains $\del_i \del^i X^+=0$. So the solution  is a free massless field and we will write it as
\begin{align}
\calX^+ (\sp, \sm)&= \calX^+_L(\sp) + \calX^+_R(\sm)\comma \\
\calXpl (\sp)&= {x^+ \over 2} + \ls^2 \pp \sp + i\ls \sum_{n \ne 0}{1\over n} 
\altil^+_n e^{-in\sp}\comma \\
\calXpr (\sm)&= {x^+ \over 2} + \ls^2 \pp \sm + i\ls \sum_{n \ne 0}{1\over n} 
\al^+_n e^{-in\sm}\period
\end{align}
Here and hereafter, we  use the calligraphic letters, such as $\calX^+$, to denote 
the fields satisfying the equations of motion. The original notation, like $X^+$, 
refers to the one without such a requirement. This distinction will be very useful and 
important, especially in the sebsequent sections. 

Next consider the equation of motion for $X^I$. Varying $\calL$ with respect to 
$X^I$, we get
\begin{align}
\del_+\del_- X^I + \mu^2 (\del_+\calXpl \del_- \calXpr) X^I =0 \period
\label{eqmXI}
\end{align}
The general solutions of this equation, $2\pi$-periodic in $\sigma$, were  given in the appendix 
 of \cite{Chizaki:2006pq}. Let us quickly reproduce them.  First make a change of 
variables and define the derivatives with respect to the new variables as  follows:
\begin{align}
(\sp, \sm) &\rightarrow (\rho_+, \rho_-) \equiv (\calXpl(\sp), \calXpr(\sm)) \comma \\
\deltil_\pm &\equiv {\del \over \del \rho_\pm} = (\del_\pm \rho_\pm)^{-1}\del_\pm
\period
\end{align}
This produces the same effect as  going to the light-cone gauge and 
 the equation for $X^I$ simplifies to 
\begin{align}
\deltil_+ \deltil_- X^I + \mu^2 X^I =0 \period \label{eqmXIrho}
\end{align}
Further form the following combinations\footnote{Here and throughout, we  take 
$\pp$  to be non-vanishing. In the Green-Schwarz formulation, 
such a restriction is   necessary  whenever we make an explicit separation of the first 
and the second class constraints.}:
\begin{align}
\ttil &\equiv {1\over 2\ls^2\pp} (\rho_++ \rho_-) \comma 
\qquad \sigtil \equiv {1\over 2\ls^2\pp} (\rho_+ - \rho_-) \period
\label{deftsigtil}
\end{align}
Under the shift $\sigma \rightarrow \sigma + 2\pi$, $\sigtil$ undergoes the same shift 
$\sigtil \rightarrow \sigtil + 2\pi$,  while $\ttil$ is invariant. In terms of these variables, 
the equation (\ref{eqmXIrho}) becomes 
\begin{align}
\left( {\del^2 \over \del \ttil^2} - {\del^2 \over \del \sigtil^2} 
\right) X^I + M^2 X^I =0 \comma \label{eqmXItil}
\end{align}
where the dimensionless ``mass" $M$ is defined  as
\begin{align}
M \equiv \alp \pp\mu = 2\ls^2 \pp \mu \period \label{defM}
\end{align}
This  is nothing but the equation of motion for a free massive field and 
the general  solution, $2\pi$-periodic in $\sig$,  can be easily obtained. 
In terms of the original variables, it can be written in  the following form:
\begin{align}
\calX^I &= \sum_n (a^I_n u_n + \atil^I_n \util_n ) \comma \\
u_n &= e^{-i (\omega_n \ttil +n \sigtil)} =  e^{-i (\lamp_n \calXpr + \lamm_n \calXpl)} \comma \label{defu} \\
\util_n &=e^{-i (\omegatil_n \ttil -n\sigtil)} =  e^{-i (\lamtilm_n \calXpr + \lamtilp_n \calXpl)} \period
\label{defutil}
\end{align}
Here  $a^I_n$ and $\atil^I_n$ are constant coefficients  and $\lam^\pm_n$ etc. are given by 
\begin{align}
\lam^\pm_n &= {1\over 2\ls^2 \pp} (\omega_n \pm n)\comma \quad 
\lamtil^\pm_n = {1\over 2\ls^2 \pp} (\omegatil_n \pm n) \comma  \label{deflam}\\
\omega_n &= \omegatil_n = {n \over |n|} \sqrt{n^2 + M^2} \qquad \mbox{for $n \ne 0$} 
\comma \label{defomnz}\\
\omega_0 &= -\omegatil_0 = M \period \label{defomz}
\end{align}
Note that for $M\ne 0$  the functions $u_n$ and $\util_n$ inherently 
depend on both $\sm$ and $\sp$ and hence it is hard to construct a purely right- or left-moving
 field out of $\calX^I$. 

The equations of motion for the fermions $\th^A$ can be derived and solved in a similar way. 
 In terms of the $\rho_\pm$ variables they become
\begin{align}
\deltil_+ \th^1&= -  \mu  \th^2  \comma \qquad 
\deltil_-\th^2=  \mu \th^1  \period \label{eqmth}
\end{align}
Combining them, $\th^A$ satisfies the same equation as $X^I$ (\ref{eqmXIrho}), namely
\begin{align}
\deltil_+ \deltil_- \th^A + \mu^2 \th^A =0\period \label{eqmthA}
\end{align}
Therefore the solution can be written in terms of the $u_n$ and $\util_n$ functions:
\begin{align}
\vartheta^A &= \sum_n ( b^A_n u_n + \btil_n^A \util_n) \period
\end{align}
Putting this back into (\ref{eqmth}), the coefficients get 
 related as
\begin{align}
\mu b^2_n &= i \lamp_n b^1_n \comma \qquad \mu \btil^2_n = i \lamtilm_n \btil^1_n 
\period
\end{align}
So only the half of these coefficients are independent. 

Finally, let us consider the equation of motion for $X^-$, which is obtained by 
 varying $\calL$  with respect to $X^+$. After some simplification using the 
 equations of motion for $\th^A$, it can be written in the $\rho_\pm$ coordinates as
\begin{align}
\deltil_+\deltil_- X^- &= \mu^2 \calX^I (\deltil_+ +\deltil_-) \calX^I 
+ i\sqrt{2}\mu (\varth^1  \deltil_+\varth^2 
- \varth^2 \deltil_-\varth^1  ) \period \label{eqmXm}
\end{align}
The RHS consists of already known functions and the LHS is the Laplacian acting on 
$X^-$. Therefore 
 $X^-$ can be readily solved  in terms of the other 
 fields once we define  a suitable inverse of the Laplacian.  As we will not need the 
resultant rather complicated expression, we do  not exhibit it. Obviously, we can always add 
a  massless free field satisfying the homogeneous part of the equation,
 which is equivalent to $\del_+\del_- X^-=0$. It is important to note that this free part
 can contain the modes of $\calX^I$ and $\varth^A$ and can only be fixed by 
requiring the correct canonical commutation relations with all the other fields. 

So we have obtained the general classical solutions of the system. They are  expressed 
 in terms of the basis functions  $u_n$ and $\util_n$, which themselves depend on the 
 modes of the field $\calX^+$. Moreover, they depend both on $\sp$ and $\sm$ 
 inseparably, as already emphasized. We will now investigate how this situation is 
 compatible with the existence of the right- and left-moving components of the 
energy-momentum tensor. 
\subsubsection{Energy-momentum tensor}
Because the Lagrangian in the SLC gauge is classically conformally invariant, 
the $++$ and the $--$ components of the energy-momentum tensor, to be denoted 
 by  $\calT_\pm$,  become the Virasoro constraints. Through a  standard procedure  one obtains 
\begin{align}
{\calT_\pm  \over T}&= 
\half \del_\pm  \Xp \del_\pm \Xm +  {1\over 4} (\del_\pm  X_I)^2  
 -{i \over \sqrt{2}} \del_\pm \Xp (\th^1 \del_\pm \th^1 + \th^2 \del_\pm \th^2)
 \nn\\
& -{1\over 4} (\del_\pm \Xp)^2 ( \mu^2 X_I^2 + 4\sqrt{2} i \mu \th^1\th^2) 
\label{calTpm}\\
&= (\del_\pm \rho_\pm)^2 \biggl[ \half \deltil_\pm  \Xp \deltil_\pm \Xm +  {1\over 4} (\deltil_\pm  X_I)^2  
 -{i \over \sqrt{2}} \deltil_\pm \Xp (\th^1 \deltil_\pm \th^1 + \th^2 \deltil_\pm \th^2)\nn\\
& -{1\over 4} (\deltil_\pm \Xp)^2 ( \mu^2 X_I^2 + 4\sqrt{2}  i \mu \th^1\th^2)
\biggr] \comma 
\end{align}
where we have exhibited the form in the $\rho_\pm$ basis as well. 

Now we  substitute the equations of motion to see if $\calT_\pm$ are functions 
 of $\sig_\pm$ respectively. Let us focus  on $\calT_+$.  Using 
$\deltil_+ \calX^+=1$ and $\mu \varth^2=-\deltil_+\varth^1$,  it  reduces to 
\begin{align}
\calT_+  &={T\over 2} (\del_+ \rho_+)^2 \biggl[  \deltil_+ \calX^- 
+  \half \left(  (\deltil_+ \calX_I)^2  -\mu^2\calX_I^2\right) 
  -i \sqrt{2}  (\varth^2 \deltil_+ \varth^2 - \varth^1 \deltil_+ \varth^1)\biggr] 
\period
\label{calTp}
\end{align}
Now let us act $\deltil_-$ on the second and the third term in the square bracket. 
By using the 
 equations of motion for $\calX_I$ and $\varth^A$, it is straightforward to get
\begin{align}
&\deltil_- \left[\half \left(  (\deltil_+ \calX_I)^2  -\mu^2\calX_I^2\right) \right]
= -\mu^2 \calX^I (\deltil_+ +\deltil_-) \calX^I  \comma \\
& \deltil_- \left[ -i \sqrt{2}  (\varth^2 \deltil_+ \varth^2 - \varth^1 \deltil_+ \varth^1)
\right] = -i\sqrt{2}\mu (\varth^1  \deltil_+\varth^2 
- \varth^2 \deltil_-\varth^1  )\period
\end{align}
Note that the expressions on the RHS are precisely those that appear in the 
 equation of motion for $X^-$ (\ref{eqmXm}), with the signs reversed. This means 
 that once-integrated equation of motion for $X^-$ is 
\begin{align}
\deltil_+ \calX^-  &=  -\half \left(  (\deltil_+ \calX_I)^2  -\mu^2\calX_I^2\right) 
  +i \sqrt{2}  (\varth^2 \deltil_+ \varth^2 - \varth^1 \deltil_+ \varth^1)
+  f_+(\sp) \comma 
\end{align}
where $f_+(\sp) $ is an arbitrary function of $\sp$. Substituting this into (\ref{calTp}), 
 $\calT_+$ collapses to 
\begin{align}
\calT_+  &={T\over 2} (\del_+ \calX^+_L)^2 f_+(\sp) \period \label{colTp}
\end{align}
In an entirely similar manner, we get
\begin{align}
\deltil_- \calX^-  &=  -\half \left(  (\deltil_- \calX_I)^2  -\mu^2\calX_I^2\right) 
  -i \sqrt{2}  (\varth^2 \deltil_- \varth^2 - \varth^1 \deltil_- \varth^1)
+  f_-(\sm) \comma \\
\calT_-  &={T\over 2} (\del_- \calX^+_R)^2 f_-(\sm) \period 
\end{align}
Thus we have a very unusual situation. Although $\calT_\pm$ are indeed functions 
 of $\sig_\pm$ respectively, explicit dependence on the fields other than $\calX^+$ 
is yet undetermined  at this stage and hence invisible. Moreover, since $\calX^I$ and 
$\varth^A$ consist of $u_n$ and $\util_n$ functions, it is not possible to construct 
 a purely left-moving or right-moving field  out of a local product  of these fields. The only
 possible way for these fields to contribute to $f_\pm(\sig_\pm)$ is through the 
coordinate-independent coefficients $a^I_n, \atil^I_n$, etc. 

 This is in striking contrast to the flat background case, 
for which $\mu$ is set to $0$ right from the beginning. In that  case, different  fields 
are mutually independent and one obtains the familiar form of $\calT_\pm$ for 
 free massless fields. This shows that the $\mu\rightarrow 0$ limit is not smooth: No matter 
 how small $\mu$ is, as long as it is finite the interactions connect up all the fields and lead 
 to the unconventional result above. 

This does not mean, however,  that the system is inconsistent in this conformally
 invariant gauge. 
It only indicates that the conformal structure in a  system where the left- and right-moving degrees of freedom are  coupled is indeed quite subtle. What we need to do is to go to the next 
 stage of the canonical analysis, namely to  set up of the canonical Poisson-Dirac  brackets 
for the fields,  and try to find  $f_\pm(\sig_\pm)$ functions which realize the correct commutation 
 relations among $X^-$ and other fields. 
\subsubsection{Poisson-Dirac  brackets for the fields and the modes}
Let us now set up the Poisson-Dirac bracket between the basic fields and their conjugates. 
We will denote the momenta conjugate to $(X^+, X^-, X^I)$ as 
$(P^-, P^+, P^I)$. They are given by  
\begin{align}
P^+ &= T \del_0 X^+ \comma \\
P^- &= T \bigl[\del_0 X^- - \del_0 X^+ ( \mu^2 X_I^2 + 4\sqrt{2} i \mu \th^1\th^2) \nn\\
& -2\sqrt{2} i (\th^1\del_+ \th^1 +\th^2\del_+\th^2)\bigr] \comma \\
P^I &= T \del_0 X^I\period 
\end{align}
As for the fermionic fields, the momenta $p^A$ conjugate to $\th^A$ take the form
\begin{align}
p^1 &=i\sqrt{2}T (\del_0X^+ -\del_1 X^+) \th^1 
= i\piplus{1}\th^1 \comma \\
p^2 &= i\sqrt{2}T  (\del_0X^++\del_1 X^+) \th^2 
= i\piplus{2} \th^2 \comma 
\end{align}
where 
\begin{align}
\piplus{1}&\equiv \sqrt{2} (P^+-T\del_1 X^+)
\comma \qquad \piplus{2} \equiv \sqrt{2} (P^++T\del_1 X^+) \period
\end{align}
Just as in the flat background case, these equations actually give the constraints
\begin{align}
d^A &\equiv p^A - i\piplus{A}\th^A =0 \period
\end{align}
They simply say that $p^A$ can be solved in terms of $\th^A$ and hence they are
of second class. 

We define the Poisson brackets  as 
\begin{align}
\Pcom{X^I(\sig,t)}{P^J(\sigp,t)} &= \delta^{IJ} \deltassp \comma 
\label{pcomXIPJ} \\
\Pcom{X^\pm(\sig, t)}{P^\mp(\sigp, t)} &= \deltassp \comma
\label{pcomXpmPmp} \\
\Pcom{\th^A_a(\sig,t)}{p^B_b(\sigp,t)} &= -\delta^{AB} \delta_{ab}
\deltassp \comma \\
\mbox{rest} &=0 \period 
\end{align}
Under this bracket, the fermionic constraints $d^A_a$ form the second class algebra
\begin{align}
\Pcom{d^A_a(\sig,t)}{d^B_b(\sigp,t)} &= 2 i \delta^{AB} \delta_{ab} 
\piplus{A}(\sig,t) \deltassp\period
\end{align}
Defining   the Dirac bracket in the standard way, $\th^A$'s  become self-conjugate 
 and satisfy\footnote{The inverse $1/\pi^{+A}$ is well-defined since its zero mode 
 $\propto \pp$ is non-vanishing.}
\begin{align}
\Dcom{\th^A_a(\sig,t)}{\th^B_b(\sigp,t)} &= {i \delta^{AB} \delta_{ab}
\over 2 \piplus{A}(\sig,t)}\deltassp \period
\end{align}
By going to the Dirac bracket,  the relations (\ref{pcomXIPJ}) and 
(\ref{pcomXpmPmp}) conitinue to hold, but the brackets between $(X^-, P^-)$ and $\th^A$, which vanished under Poisson,  become non-trivial\footnote{The bracket $\Dcom{X^-}{P^-}$
 still vanishes due to $\th^A_a \th^A_a =0$.}. One finds
\begin{align}
\Dcom{X^-(\sig,t)}{\th^A(\sigp,t)} &= -{1\over \sqrt{2} \piplus{A}(\sig,t)}
 \th^A 
\deltassp \comma \\
\Dcom{P^-(\sig,t)}{\th^A(\sigp,t)} &= -{1\over \sqrt{2} \piplus{A}(\sigp,t)} \th^A (\sigp,t)
\deltapssp \period
\end{align}
However, if we define the combination $\Theta^A_a \equiv \sqrt{2\piplus{A}} \th^A_a$, 
it is not difficult to check that 
they satisfy 
\begin{align}
\Dcom{\Theta^A_a(\sig,t)}{\Theta^B_b(\sigp,t)} &= i\delta^{AB} \delta_{ab}
\deltassp \comma 
\end{align}
and commute with all the other fields. So the fields to used 
 are $(X^\pm, X^I, P^\pm, P^I, 
\Theta^A_a)$, which satisfy the canonical form of the Diract bracket relations. 

We now  come to the question of finding the brackets  among 
 the modes  $a^I_n, \atil^I_n, b^A_n, \btil^A_n$,  etc. which realize these 
canonical equal-time Dirac bracket relations  for the fields. In the case of  free fields, 
this is a textbook matter  as one can easily express the modes in terms of the fields
using the completeness relations of the basis functions $e^{in\sig_\pm}$ at each  time slice. 
Here we encounter a serious difficulty: Our basis functions, $u_n$ and $\util_n$
 obtained  in (\ref{defu}) and (\ref{defutil}), are highly  complicated functions  of $\sig$ at equal $t$,  since $\calX^+_L$ and $\calX^+_R$,  appearing in the exponent,  themselves depend on $\sig$ exponentially. Although we have been able to  express the modes, 
such as $a^I_n, \atil^I_n$,  in terms of fields, but  such expressions turned out to be
 quite formal and complicated,  and so far not of practical use. 

In this regard, note that if $\ttil(t, \sig)$ and $\sigtil(t,\sig)$, 
 defined in (\ref{deftsigtil}),  were the time and the space variables, the situation 
would have been  much simpler, just like in the light-cone gauge. 
Thus, the difficulty we encountered  is  due to the non-trivial relation between the 
symplectic structures in the canonical $(t,\sig)$ basis and the $(\ttil, \sigtil)$ basis, 
 which are connected by a field-dependent conformal transformation.

Fortunately, there is a  nice way out of this problem. As we will explain  in the next subsection, 
we can formulate the theory, including its dynamics, entirely in terms of the Fourier modes 
 of the phase-space fields at $t=0$, without the use of the solutions of the equations 
 of motion. 
\subsection{Phase-space formulation and the Virasoro algebra}
\subsubsection{Basic idea}
In ordinary field theories, the knowledge of the Poisson(-Dirac) brackets of the phase-space
 fields at one time is not enough to describe the dynamics. Although one can promote  these 
brackets into quantum brackets,  it is not possible to compute the correlation functions 
 of the  fields at different times. This is why one needs to first obtain the solutions 
 of the equations of motion and then  find the brackets for the $t$-independent modes in them 
in order  to construct the quantized fields at an arbitrary time. 

The situation can be  different, however, for a theory in which  {\it the  Hamiltonian  is a member 
of the generators of a  large  symmetry algebra}. In such a case, provided that the symmetry 
 is powerful enough, the representation theory of the algebra in the field space 
alone may  fix the dynamics 
as well. String theories  in a conformally invariant gauge belong to this 
 category.  To our knowledge, this observation has not been duly utilized in the past. 
This is simply because it has not been needed: Solvable string theories have been
 limited in number  and the usual canonical procedure was sufficient to quantize
 and solve them. We would like to emphasize that the method to be described below
 is very powerful for the  cases where the equations of motion are hard to solve due to 
non-linearity  or, as in our case, the canonical quantization is difficult. 
\subsubsection{Classical Viraosoro algebra in the phase space}
Hereafter we will be dealing with the phase-space fields, which are not subject to any 
 equations of motion.  The central objects are  the energy-momentum tensors $\calT_\pm$, 
which are  given in (\ref{calTpm}).  To express them in terms of the phase-space variables, 
 it is convenient to introduce the following dimensionless fields $\{A, B, \Pitil, \Pi, S \}$, 
 with appropriate sub- and super-scripts, 
 and a dimensionless constant $\muhat$:
\begin{align}
X &= {1\over \sqrt{2\pi T}} A \comma \qquad P = \sqrt{T \over 2\pi} B  
\comma \nn\\
\Pitil &= {1\over \sqrt{2}} (B +\del_1 A) \comma \qquad \Pi 
 = {1\over \sqrt{2}} (B -\del_1 A) \comma \nn\\
\Th &= -{i\over \sqrt{2\pi}} S  \comma \qquad 
\muhat = {\mu\over \sqrt{2\pi T}}  \period 
\end{align}
Then, $\calT_\pm$ can be written as 
\begin{align}
\calT_+ &= \half (\calH + \calP) \nn\\
&= {1\over 2\pi} \biggl(\Pitil^+\Pitil^- + \half \Pitil_I^2 +{i \over 2} S^2 \del_1 S^2 
 + {\muhat^2 \over 2} \Pitil^+ \Pi^+ A_I^2 -{i\muhat \over \sqrt{2}}
\sqrt{\Pitil^+\Pi^+} S^1 S^2  \biggr) \comma \label{calTplus}\\
\calT_- &= \half (\calH - \calP) \nn\\
&={1\over 2\pi} \biggl( \Pi^+\Pi^- + \half \Pi_I^2 -{i \over 2} S^1 \del_1 S^1 
 + {\muhat^2 \over 2} \Pitil^+ \Pi^+ A_I^2 - {i\muhat \over \sqrt{2}}
\sqrt{\Pitil^+\Pi^+} S^1 S^2 \biggr)\period \label{calTminus}
\end{align}
\nullify{
\begin{align}
\calT_\pm &= \half (\calH \pm \calP)\comma \\
\calP &= P^+ \del_1 X^- + P^- \del_1 X^+ + P^I \del_1 X_I 
 -{i \over 2} \Theta^1 \del_1 \Theta^1 - {i \over 2}  \Theta^2 \del_1 \Theta^2 \comma \label{calP}\\
\calH &= {P^- P^+ \over T} + T \del_1 X^+ \del_1 X^- 
+ \half {P_I^2 \over T} + {T \over 2} (\del_1 X_I)^2 
+ {\mu^2 \over 2T} X_I^2 \left( (P^+)^2 -(T\del_1 X^+)^2 \right)  \nn\\
& +{i \over 2} \Theta^1 \del_1 \Theta^1 - {i \over 2}  \Theta^2 \del_1 \Theta^2 
+ {i \mu \over T } \sqrt{(P^+ +T \del_1 X^+) (P^+ -T \del_1 X^+) } \,
 \Theta^1  \Theta^2
\period \label{calH}
\end{align}
}
$\calP$ and $\calH$ are, respectively, the momentum density and the Hamiltonian density. 
Since the basic fields satisfy the canonical form of the Dirac bracket relations  given 
 in the previous subsection, we can compute the brackets among $\calP$ and $\calH$ and
 hence among $\calT_\pm$. We need to make use of the following formulas
for the derivatives of the $\delta$-function  for a general field $\calO$:
\begin{align}
\calO(\sigp)\deltapssp &= \calO(\sig)\deltapssp + \del_1 \calO(\sig) \deltassp  \comma \\
\calO(\sigp)\delta''(\sig-\sigp) &= \calO(\sig)\delta''(\sig-\sigp) + 2\del_1 \calO(\sig)
\deltapssp + \del_1^2 \calO(\sig)\deltassp \comma 
\end{align}
which must be understood in the sense of distributions. After  a straightforward but 
long calculation, we obtain the expected results:
\begin{align}
\Dcom{\calH(\sig,t)}{\calH(\sigp,t)} &= \Dcom{\calP(\sig,t)}{\calP(\sigp,t)} \nn\\
& = 2\calP(\sig,t) \deltapssp + \del_1 \calP(\sig,t) \deltassp \comma \\
\Dcom{\calP(\sig,t)}{\calH(\sigp,t)} &= \Dcom{\calH(\sig,t)}{\calP(\sigp,t)}   \nn\\
& =2\calH(\sig,t) \deltapssp + \del_1 \calH(\sig,t) \deltassp  \comma \\
\Dcom{\calT_\pm(\sig,t)}{\calT_\pm(\sigp,t)} 
&= \pm 2 \calT_\pm(\sig,t)\deltapssp \pm 
\del_1 \calT_\pm(\sig,t) \deltassp\comma  \label{DcomcalT}\\
\Dcom{\calT_\pm(\sig,t)}{\calT_\mp(\sigp,t)} &=0 \period
\end{align}
The last two lines show that $\calT_\pm$ form mutually commuting  Virasoro algebras\footnote{$\pm$ signs on the RHS are just right for the Virasoro mode operators $T^\pm_n$ to satisfy the same algebra. See the discussion below.}. 
Integrating the first two equations with respect to  $\sigp$ and identifying $\int d\sigp \calH(\sigp,t)$ to be  the Hamiltonian $H$, which generates the time-development 
of a field $A(\sig,t)$  as $\del_0 A = \Dcom{A}{H}$, one readily finds
\begin{align}
\del_0\calH &= \Pcom{\calH}{H} = \del_1 \calP \comma \qquad 
\del_0\calP = \Pcom{\calP}{H} = \del_1 \calH \period
\end{align}
Combining them, we get $\del_\mp \calT_\pm =0$,  showing that  $\calT_\pm= \calT_{\pm}(\sigma_\pm)$. Therefore, we can define the Virasoro mode operators 
$T^\pm_n$ by 
\begin{align}
\calT_\pm &= {1\over 2\pi} \sum_n T^\pm_n e^{-in(t \pm \sig)} \period
\end{align}
Putting this into (\ref{DcomcalT}), one verifies that $T^\pm_n$  satisfy the usual 
 form of the classical Virasoro algebra, namely
\begin{align}
\Dcom{T^\pm_m}{T^\pm_n} &= {1\over i} (m-n) T^\pm_{m+n} 
\comma \qquad \Dcom{T^\pm_m}{T^\mp_n} =0 \period
\end{align}
What is important here is that since $T^\pm_n$ are independent of $t$ and $\sigma$ 
 they can be obtained from $\calT_\pm$ at one time slice, which we take to be $t=0$:
\begin{align}
T^\pm_n &= \int_0^{2\pi}  d\sig e^{\pm in\sig} \calT_{\pm}(\sig, t=0) \period
\label{defTpmn}
\end{align}
But at $t=0$, we know the exact Dirac brackets for the fields composing $\calT_\pm$ and hence we can quantize them in the standard way. Moreover, any properties of the
 system which are dictated by $T^\pm_n$ can in principle be calculable. 
The spectrum of physical states is one such quantity and in Sec.4 
 we will demonstrate  that it can  indeed be computed. 

The vertex operators that 
 create these physical states sould also be obtainable. Once they are constructed, 
 one can calculate the correlation functions at unequal times, which reflect 
the dynamics of the system. This and the related matters will be 
investigated in a separate publication. 

\section{Quantization and quantum Virasoro operators}
\subsection{Quantization of basic fields}
Let us now quantize the phase-space fields by replacing the Dirac brackets by 
the (anti-)commutators and $\deltassp \rightarrow i \deltassp$. 
As already said, we will do this  at time $t=0$ and hence drop 
$t$ for all the fields from now on. 
The non-vanishing (anti-)commutators among the basic fields are then given by 
\begin{align}
\com{A^+(\sig)}{B^-(\sigp)} &= 
\com{A^-(\sig)}{B^+(\sigp)} =2\pi i \deltassp \comma  \\
\com{A^I(\sig)}{B^J(\sigp)} &= 2\pi i \delta^{IJ} \deltassp \comma \\
\acom{S^A_a(\sig)}{S^B_b(\sigp)} &= 2\pi \delta^{AB}\delta_{ab} \deltassp \period
\end{align}
We take the Fourier mode expansions  of $A^\star, B^\star$ ($\star = (\pm, I)$) 
and $S^A_a$ to be 
\begin{align}
A^\star(\sig) &= \sum_n A^\star_n e^{-in\sig} \comma \qquad B^\star(\sig) = 
\sum_n B^\star_n e^{-in\sig} \comma \qquad S^A_a(\sig) = \sum_n S^A_{a,n} e^{-in\sig} \period
\end{align}
Then, these modes satisfy the simple (anti-)commutation relations:
\begin{align}
\com{A^\pm_m}{B^\mp_n} &= i \delta_{m+n,0} \comma \quad 
\com{A^I_m}{B^J_n} = i \delta^{IJ} \delta_{m+n,0} \comma  \\
\acom{S^A_{a,m}}{S^B_{b,n}} &= \delta^{AB}\delta_{ab} \delta_{m+n,0}
\comma \qquad \mbox{rest=0} \period 
\end{align}
As for the $\Pitil$ and $\Pi$ fields, we have 
\begin{align}
\com{\Pitil^+(\sig)}{\Pitil^-(\sigp)} &=  
-\com{\Pi^+(\sig)}{\Pi^-(\sigp)} =2\pi i \deltapssp   \comma \nn\\
\com{\Pitil^I(\sig)}{\Pitil^J(\sigp)} &= -\com{\Pi^I(\sig)}{\Pi^J(\sigp)} =
2\pi i \delta^{IJ} \deltapssp \comma \nn\\
\com{\Pitil^I(\sig)}{A^J(\sigp)} &= \com{\Pi^I(\sig)}{A^J(\sigp)}
= -{i \over \sqrt{2}} \delta^{IJ} \deltassp \period
\end{align}
From the definition, their modes are given by 
\begin{align}
\Pi^\pm_n &\equiv {1\over \sqrt{2}}(B^\pm_n +inA^\pm_n) \comma \quad 
\Pi^I_n \equiv {1\over \sqrt{2}}(B^I_n +inA^I_n) \comma \\
\Pitil^\pm_n &\equiv {1\over \sqrt{2}}(B^\pm_n -inA^\pm_n) \comma \quad 
\Pitil^I_n \equiv {1\over \sqrt{2}}(B^I_n -inA^I_n) \comma 
\end{align}
and they satisfy the following commutation relations:
\begin{align}
\com{\Pi^\pm_m}{\Pi^\mp_n} &= -m \delta_{m+n,0}
\comma \quad \com{\Pi^I_m}{\Pi^J_n} = -m \delta^{IJ} \delta_{m+n,0}
\comma \label{ComPi}\\
\com{\Pitil^\pm_m}{\Pitil^\mp_n} &= m \delta_{m+n,0}
\comma \quad \com{\Pitil^I_m}{\Pitil^J_n} = m \delta^{IJ} \delta_{m+n,0}
\period
\end{align}
Note the  difference in sign for $\Pi$ commutators and $\Pitil$ commutators.

Let us make a remark. Although these quantum fields satisfy the free-field commutation
 relations, they are not bonafide  free fields. The reason is that 
they time-develop non-trivially according to the Hamiltonian 
$H = \int d\sig \calH$, which contains non-quadratic  terms. Nevertheless, one can compute
 all the commutators  as if they were free massless fields. This is a great 
 advantage of the present formalism. 
\subsection{Normal-ordering for quantum Virasoro operators}
Now we want to define the quantum Virasoro operators with an appropriate normal-ordering
and see if they form proper quantum Virasoro algebras. 
This turned out to be a rather delicate problem due precisely to the terms depending 
 on the ``mass" parameter $\muhat$. 

To begin, let us recall the classical Virasoro algebra derived in (\ref{DcomcalT}). 
It tells us  that both $-\calT_-(\sig)$ and $\calT_-(-\sig)$ satisfy the same algebra as $\calT_+(\sig)$  including the signs. In terms of modes, it  means
that $-T^-_{-n}$ and $T^-_n$ satisfy the same standard form of the 
Virasoro algebra as $T^+_n$. Accordingly, we will study two different quantum extensions. 
\subsubsection{Phase-space normal-ordering}
First, consider the case where we take the two sets of Virasoro operators to be 
\begin{align}
\calL_+ (\sig)&\equiv \calT_+(\sig) =
{1\over 2\pi} \sum_n L^+_n e^{-in\sig} \comma \qquad L^+_n \equiv T^+_n
\comma \\
\calL_- (\sig)&\equiv -\calT_-(\sig) = {1\over 2\pi} 
\sum_n L^-_n e^{-in\sig}\comma \qquad 
L^-_n \equiv -T^-_{-n} \period \label{defLpmn}
\end{align}
The most natural normal-ordering scheme in this case is to regard 
$B^\star_n (n \ge 0), 
A^\star_n (n \ge 1), S^A_{a,n} (n \ge 1)$ as ``annhiliation operators", 
where $\star = (\pm, I)$. We will call it the {\it phase-space normal-ordering}. 
Its naturalness  can be seen, for instance,  by quickly computing the central 
charge terms  for the bosons in ``$\pm$" sectors. As a matter of fact, since 
 we know that $\calL_\pm$ form the standard Virasoro algebras  classically, what we have to 
 study  are the  quantum contributions, such as the central charge terms, 
 coming from  the ``double-contractions".  

Consider first the commutator $\com{\calL_+(\sig)}{\calL_+(\sigp)}$, which is defined 
 as $\calL_+(\sig-i\ep) \calL_+(\sigp) - \calL_+(\sigp-i\ep) \calL_+(\sig)$, with 
 an infinitesimal positive regulator $\ep$. 
There are two  different types of  double contractions in this computation. 
One type consists of the usual contributions which are present for $\muhat=0$ case as well.
They produce the familiar 
c-number anomaly of the form $-(i/24\pi)(\delta'''(\sig-\sigp) -\deltapssp)$
 for a boson and $(-i/24\pi) (\half \delta'''(\sig-\sigp) +\deltapssp)$ for a self-conjugate 
 periodic fermion.  The other type, which will be our central focus from now on,  consists of the contributions from the following 
 commutators in the bosonic and the fermionic sectors respectively:
\begin{align}
C_B &= {1\over (2\pi)^2} \left(\com{\half \Pitil_I^2(\sig)}{{\muhat^2 \over 2}\Pitil^+\Pi^+ A_I^2(\sigp)}
-(\sig \leftrightarrow \sigp) \right)\comma \\
C_F &=  {1\over (2\pi)^2}  \com{{-i\muhat\over \sqrt{2}}\sqrt{\Pitil^+\Pi^+}
S^1S^2(\sig)}{{-i\muhat\over \sqrt{2}}\sqrt{\Pitil^+\Pi^+}S^1S^2(\sigp)} 
\period \label{CF}
\end{align}
As shown in Appendix A.1, they yield the same non-vanishing singularity  but with 
 opposite signs, namely 
\begin{align}
C_B &= -C_F = -{i \muhat^2 \over \pi} ( 2 \Pitil^+\Pi^+ \deltapssp
+ \del_\sig(\Pitil^+\Pi^+) \deltassp ) \comma \label{CBCF}
\end{align}
and hence cancel each other precisely. Similar cancellations take place also 
 in $\com{\calL_-(\sig)}{\calL_-(\sigp)}$ and $\com{\calL_+(\sig)}{\calL_-(\sigp)}$. 
In this way we obtain the closed Virasoro algebra
\begin{align}
\com{\calL_\pm(\sig)}{\calL_\pm(\sigp)} &= i \biggl( 2\calL_\pm(\sig) \deltapssp 
 + \del_\sig \calL_\pm(\sig) \deltassp \nn\\
& -{1\over 24\pi} (14 \delta'''(\sig-\sigp) -2\deltapssp) \biggr) \comma 
\label{ComLpmLpm}\\
\com{\calL_+(\sig)}{\calL_-(\sigp)} &= 0 \period
\end{align}
It should be emphasized  that the bosonic string in the plane-wave  background 
 is  conformally invariant classically but not quantum-mechanically: The contribution
 from the fermions  coupled to the RR flux is crucial in cancelling the potential operator anomaly (\ref{CBCF}). 

Now we have to discuss the issue of the central charge. As it stands, the central charge 
 is only 14, 10 from the bosons and 4 from the self-conjugate fermions. We need to supply 
 12 more units to construct  a consistent string theory. This situation, however, is exactly the 
 same as for  the flat background in the SLC gauge and the proper 
cure is known\cite{Berkovits:2004tw}. One only needs to add the 
quantum corrections of the form 
\begin{align}
\Delta \calL_+ = -{1\over 2\pi} \del_\sig^2 \ln \Pitil^+\comma  
\qquad \Delta \calL_- = {1\over 2\pi} \del_\sig^2 \ln \Pi^+ \comma 
\end{align}
to $\calL_\pm$, respectively. It is straightforward to verify 
\begin{align}
&\com{\calL_\pm (\sig)}{\Delta \calL_\pm (\sigp)} + 
\com{\Delta \calL_\pm (\sig)}{\calL_\pm (\sigp)} \nn\\
& = i \biggl( 2 \Delta \calL_\pm (\sig) \deltapssp + \del_\sig (\Delta \calL_\pm (\sig)) \deltassp
-{1\over 24\pi} 12 \delta'''(\sig-\sigp) \biggr) \period
\end{align}
This means that $\Delta \calL_\pm$ are primary operators of dimension 2,  except that 
 they provide the desired  12 units of central charge. Hereafter,  $\calL_\pm$ will be understood
 to include these quantum corrections. 
\subsubsection{Massless normal-ordering}
Although we have already found  that the phase-space normal-ordering scheme works, 
 it is instructive to examine another scheme, one which appears natural  for the 
 case with $\muhat =0$, \ie for the usual free massless fields in 2 dimensions. 

For $\muhat =0$, the terms proportional to $A_I^2$ and $S^1S^2$ are absent and, 
in particular, the bosonic sector consists of $\Pitil^\star$ and $\Pi^\star$ fields only. 
This suggests that we may define  the normal-ordering based on the modes of $\Pitil^\star$
 and $\Pi^\star$. However, as we saw in (\ref{ComPi}),  the commutators of  $\Pi^\star_n$'s are  opposite in sign to those of $\Pitil^\star_n$'s. Therefore, in the ``$-$" sector, it is 
natural to reverse the mode number and define $\Pihat^\star_n \equiv 
\Pi^\star_{-n}$. At the field level, we introduce 
\begin{align}
\Pihat^\star(\sig) &\equiv \Pi^\star(-\sig) = \sum_n \Pihat^\star_n e^{-in\sig} 
\period
\end{align}
Then, $\Pihat^\star$ statisfies the same commutation relation as $\Pitil^\star$, \ie 
$\com{\Pihat^\star(\sig)}{\Pihat^\star(\sigp)} = 2\pi i \deltapssp$. 
Correspondingly, the Virasoro generator for the ``$-$" sector should be taken as 
$\calThat_-(\sig) \equiv \calT_-(-\sig)$, which classically satisfies the standard form of 
 the Virasoro algebra, as discussed at the beginning of Sec.3.2. The fermion $S^1$ appearing 
 in the ``$-$" sector should also be treated in a similar way. Namely we define 
$\Shat^1_n \equiv S^1_{-n}$ and introduce 
$\Shat^1(\sig) \equiv S^1(-\sig) = \sum_n \Shat^1_n e^{-in\sig}$. What we shall call 
 the {\it massless normal-ordering} is then defined by regarding $\Pitil^\star_n (n\ge 0)$,
$ \Pihat^\star_n (n \ge 0)$,  $\Shat^1_n (n \ge 1)$, $S^2_n (n\ge 1)$ as ``annihiliation 
 operators" and  will be denoted by the symbol $\ant \quad \ant$. For $\muhat =0$, 
 the Virasoro operators $\calT_+(\sig)$ and $\calThat_-(\sig)$ consist of 
mutually independent sets of fields and are readily shown to satisfy the 
isomorphic quantum Virasoro algebras of the standard form. 

Let us  try to extend this scheme to the $\muhat \ne 0$ case. We now have to express the 
 non-zero modes of $A^I$  in terms of $\Pitil^I_n$ and $\Pihat^I_n$ as 
\begin{align}
A^I_n &= {i \over \sqrt{2}} {1 \over n} (\Pitil^I_n -\Pihat^I_{-n}) \period
\end{align}
The splitting of the field  $A_I(\sig)$ into the annihilation ($(+)$) and the creation(($-$))  parts 
should be made as 
\begin{align}
A_I(\sig) &= {\calA}_I\toplus(\sig) + {\calA}_I\tominus(\sig) \comma \\
{\calA}_I\toplus(\sig) &=    {i \over \sqtwo} \sum_{n\ge 1} {1\over n}
 \left( \Pitil_n e^{-in\sig} + \Pihat_n e^{in\sig} \right) \comma \\
{\calA}_I\tominus(\sig)&= A_{I,0}-{i \over \sqtwo} \sum_{n\ge 1} {1\over n}
 \left( \Pitil_{-n} e^{in\sig} + \Pihat_{-n} e^{-in\sig} \right) \period
\end{align}
Although $A_I$ commutes with itself, ${\calA}_I\toplus $ and ${\calA}_I\tominus$ 
do not, and hence $\Pitil^+\Pi^+A_I^2$  must be defined  with non-trivial normal-ordering,
 which  discards the operator  $\Pitil^+\Pi^+$ with a divergent 
 coefficient\footnote{Since, as can be easily checked, 
  $\Pitil^+\Pi^+$ is  an exactly marginal 
operator, subtraction of such an operator by itself does not interfere with conformal invariance.}. 

The Virasoro operators 
 take the form (suppressing  the normal-ordering symbol $\ant \quad \ant$) 
\begin{align}
\calT_+(\sig) &={1\over 2\pi} \biggl(\Pitil^+\Pitil^-(\sig)  + \half \Pitil_I^2 (\sig) +{i \over 2} S^2 \del_1 S^2 (\sig) \nn\\
& + {\muhat^2 \over 2} \Pitil^+ \Pi^+(\sig) A_I^2 (\sig) -{i\muhat \over \sqrt{2}}
\sqrt{\Pitil^+\Pi^+(\sig)} \Shat^1 (-\sig) S^2 (\sig)\biggr)  \comma \\
\calThat_-(\sig) &= {1\over 2\pi} \biggl(\Pihat^+\Pihat^- (\sig)+ \half \Pihat_I^2(\sig)
 +{i \over 2} \Shat^1 \del_1 \Shat^1(\sig) 
\nn\\
& + {\muhat^2 \over 2} \Pitilhat^+ \Pihat^+(\sig)  A_I^2(-\sig) 
- {i\muhat \over \sqrt{2}}
\sqrt{\Pitilhat^+\Pihat^+(\sig)} \Shat^1(\sig) S^2(-\sig) \biggr)\period
\end{align}
An unusual feature is that due to the simultaneous presence of the 
hatted and unhatted operators, 
the mode number  is not conserved for certain  terms in the Virasoro generators. 
The calculation of the commutators between the Virasoro generators above proceeds
 in the similar way as before. Again the crucial part is the computation of the double-contraction 
 contributions, which is described in Appendix A.2. 

There are several differences from 
 the phase-space normal-ordering scheme. First, the contribution
 from the fermions, corresponding to (\ref{CF}),  vanishes.  On the other hand
 the bosonic contribution  in  the commutators $\com{\calT_+(\sig)}{\calT_+(\sigp)}$ and 
 $\com{\calThat_-(\sig)}{\calThat_-(\sigp)}$ is the same as $C_B$ given in 
(\ref{CBCF}). Consequently, these Virasoro commutators contain extra uncanceled operator anomalies  of 
 the form 
\begin{align}
\com{\calT_+(\sig)}{\calT_+(\sigp)}_{extra} &
= -{i \muhat^2 \over \pi} ( 2 \Pitil^+\Pi^+ (\sig)\deltapssp
+ \del_\sig(\Pitil^+\Pi^+(\sig)) \deltassp ) \comma \\
\com{\calThat_-(\sig)}{\calThat_-(\sigp)}_{extra} 
&= -{i \muhat^2 \over \pi} ( 2 \Pitilhat^+\Pihat^+(\sig) \deltapssp
+ \del_\sig(\Pitilhat^+\Pihat^+(\sig)) \deltassp ) \period
\end{align}
Second, in the remaining commutator $\com{\calT_+(\sig)}{\calThat_-(\sigp)}$,  a similar 
 anomaly of the following form  is produced from  the bosonic sector\footnote{Due to 
 the definition $\calThat_-(\sig) = \calT_-(-\sig)$, the argument of the $\delta$-function below  is $\sig + \sigp$, not $\sig-\sigp$. In any case, the point is that the 
Virasoro generators for the ``$\pm$" sectors do not decouple.}:
\begin{align}
\com{\calT_+(\sig)}{\calThat_-(\sigp)} &= 
 -{ i\muhat^2 \over \pi} \del_\sig (\Pitil^+\Pi^+)(\sig) \delta(\sig + \sigp) \period 
\end{align}
As far as we know,  these operator anomalies cannot be removed by adding some quantum 
 correction to the Virasoro generators. 

The conclusion is that while  the massless normal-ordering works perfectly for $\muhat=0$ case, it is plagued with operator anomalies for $\muhat \ne 0$.  
\section{Spectrum of physical states}
\subsection{BRST operator and its cohomology}
With the  quantum Virasoro generators defined with the phase-space normal-ordering, we can construct the nilpotent 
BRST operators $Q$ and $\Qtil$ for the ``$-$" and the ``$+$" sectors in the usual way
 and study their cohomologies. As it will become evident, the analysis is exactly parallel to 
 the case of the free bosonic string. This is because the structure of the unphysical quartets 
 and the mechanism of their decoupling from the  physical space 
is  identical to that case,  despite the presence of the additional interaction terms and (physical) 
 fermions. Thus one may refer to  the standard argument, a particularly 
suited  one being  that given  in the Polchinski's book\cite{Polchinski:1998rq}, and 
state the result. 
However, for self-containedness  and for a need of some additional explanations, we shall 
recapitulate   the essential part of the argument. 

To make the presentation concise and transparent, it is convenient to first recall 
the following  basic   theorem on the cohomology, which will be repeatedly invoked. \\
{\it Theorem}: \quad Let $(\calQ, \calK, \calN)$ be a triple of operators, 
$\calQ, \calK$ being fermionic and $\calN$  bosonic, and assume that they satisfy the following relations:
\begin{align}
\calQ^2 &=0 \comma \qquad 
\acom{\calQ}{\calK} = \calN \period \label{QKN}
\end{align}
Then the cohomology of $\calQ$ is in the kernel $\Ker (\calN)$ of $\calN$. \\
{\it Proof:} \quad The proof is simple. Let $\ket{\Phi}$ be a $\calQ$-closed state,
 {\it i.e.}  $\calQ \ket{\Phi} =0$. Then, from the relation above, 
$(\calQ \calK +\calK\calQ)\ket{\Phi} = \calN \ket{\Phi} = \calQ \calK \ket{\Phi}$. 
Now if $\ket{\Phi}$ is {\it not} in $\Ker (\calN)$, then,  
this can be rewritten as $\ket{\Phi} = \calQ(\calN\inv \calK \ket{\Phi})$, where 
 we used the commutativity 
 of $\calQ$ and $\calN$ which follows  from  (\ref{QKN}). 
Thus, such a state is $\calQ$-exact and hence the cohomology of $\calQ$ can only be 
 in $\Ker (\calN)$. 

Now we begin the analysis for our system. Since the argument is entirely similar 
 for $Q$ and $\Qtil$, we will focus our attention on $Q$, which is given by 
\begin{align}
Q &= \sum_n c_{-n}L^-_n -\half \sum_{m,n} (m-n) :c_{-m}c_{-n} b_{m+n}:
\period \label{BRSTQ}
\end{align}
Define  the ``light-cone number" operator $N_{lc}$ by
\begin{align}
N_{lc} &\equiv \sum_{n\ge 1} {1\over n} \left( \Pi^-_{-n} \Pi^+_n 
-\Pi^+_{-n} \Pi^-_{n} \right) 
\comma 
\end{align}
which assigns $+1$ to  $\Pi^+_n$ and $-1$ to  $\Pi^-_n$ for $n\ne 0$. Together 
 with the non-zero modes of ghosts,  $\Pi^\pm_n$  will  form the unphysical quartet $q_{n} \equiv  (\Pi^\pm_{n}, c_{n}, b_{n}) $.  In terms of this grading, $Q$ is split into
\begin{align}
Q &= Q_{-1} + Q_0 + Q_{\ge 1}\comma \label{Qgrading}
\end{align}
where the subscript refers to the light-cone number. $Q_{-1}$ is given by 
\begin{align}
Q_{-1} &= -\pphat  \sum_{n \ne 0} \Pi^-_{-n} c_n \comma \qquad \pphat \equiv 
\Pi^+_0\comma 
\end{align}
while $Q_0$ is of the same form as $Q$ in (\ref{BRSTQ}) except that $\Pi^\pm$
 in  $L^-_n$ are replaced 
 by their  zero modes  $\hat{p}^\pm$. The remaining piece $Q_{\ge 1}$ 
 is complicated for our system but it contains at least one non-zero mode of $\Pi^\pm$. 
{\it The important point is that  the explicit forms of $Q_0$ and $Q_{\ge 1}$
 will not be required.}  This is the reason why we can apply   the reasoning for the free bosonic 
 string  to  our case as well. The only information needed will be the relations that follow 
 from the nilpotency $Q^2=0$ and the ghost number structure of the  states. 

Since $Q_{-1}^2=0$ follows immediately from $Q^2=0$ and the grading (\ref{Qgrading}) above, one first considers  the cohomology 
 of this simple operator $Q_{-1}$. To make use of  the basic theorem, introduce the operator
\begin{align}
K &\equiv {1\over \pphat} \sum_{n\ne 0} \Pi^+_{-n} b_n \period
\end{align}
Its anti-commutator with $Q_{-1}$ produces a bosonic operator 
\begin{align}
N_q &= \acom{K}{Q_{-1}} = \sum_{n=1}^\infty \left( n (b_{-n}c_n 
+ c_{-n}b_n)  -\Pi^-_{-n}\Pi^+_n -\Pi^+_{-n} \Pi^-_n \right)\period
\end{align}
$N_q$ counts the Virasoro level of the quartet members  in the sense 
$\com{N_q}{q_{-n}} = n q_{-n}$. Evidently, the set $(Q_{-1}, K, N_q)$ forms 
 a triple. Hence we can apply the basic theorem to  learn that the 
 cohomology of $Q_{-1}$ must be in $\Ker(N_q)$, namely the transverse Hilbert space 
 $\calH_T$  where  the quartet members are not excited. 

 In fact one can easily prove  that the $Q_{-1}$-
 cohomology is equal  to $\Ker(N_q)$. Although $\Ker(N_q)$ contains a sector with $c_0$ 
ghost, to make the presentation  shorter, we will hereafter impose an additional condition
 $b_0 \ket{\Psi} =0$, as in \cite{Polchinski:1998rq},  to eliminate such a sector\footnote{
The full analysis including the  $c_0$ sector can of course be done,  with  more involved 
 ghost number analysis.}. In this setting, the states in $\Ker(N_q)$ carry the ghost number
 of the ghost vacuum, namely $-\half$,  and conversely any state with a ghost number different from this  is  not in $\Ker(N_q)$. 

Now let us show first that all the states in $\Ker(N_q)$ are  $Q_{-1}$-closed. Let $\ket{\psi_0}$ be in $\Ker(N_q)$, {\it i.e.}  $N_q \ket{\psi_0} =0$. Apply 
$Q_{-1}$ and use the commutativity of $Q_{-1}$ and $N_q$  (which follows from 
(\ref{QKN})). Then, we get $N_q (Q_{-1} \ket{\psi_0}) =0$. But since $Q_{-1}$ carries 
  ghost number $1$, $N_q$ on $Q_{-1} \ket{\psi_0}$ is non-vanishing. 
It follows that $Q_{-1} \ket{\psi_0}=0$. 

The proof that $\ket{\psi_0}$ is not $Q_{-1}$-exact
is equally straightforward. Assume that $\ket{\psi_0} = Q_{-1} \ket{\chi}$ for some $\ket{\chi}$.
 Applying $N_q$ one gets $0 = N_q\ket{\psi_0} = Q_{-1}(N_q \ket{\chi})$, showing 
 that $N_q\ket{\chi}$ is $Q_{-1}$-closed. But since $N_q \ket{\chi}$ is not in $\Ker (N_q)$
 due to its ghost number, the basic theorem tells us that it must be of the $Q_{-1}$-exact
 form, {\it i.e.} $N_q \ket{\chi}=Q_{-1} \ket{\xi}$. As $N_q$ is invertible in this sector, 
 this means that $\ket{\chi} = Q_{-1} (N_q\inv \ket{\xi})$.  Hence $\ket{\psi_0}
 = Q_{-1} \ket{\chi}$ vanishes identically and there is no $Q_{-1}$-exact state in $\Ker(N_q)$. 

One can apply exactly the same method  to the study of $Q$-cohomology itself. 
To this end, one introduces  another triple of operators $(Q, K, N)$, 
 where $N$ is given by 
\begin{align}
N = \acom{Q}{K}  \period 
\end{align}
Following the same logic as before, 
we immediately conclude that $Q$-cohomology is isomorphic  to $\Ker(N)$. 

The final step is to prove that in fact  $\Ker(N)$ and $\Ker(N_q)$ are the same. 
This is done by an explicit construction of the isomophism between these spaces. 
For this purpose, we split the operator $N$ into two parts as follows:
\begin{align}
N &= N_q + \Ncheck, \comma \qquad \Ncheck = \acom{Q_0 + Q_{\ge 1}}{K} 
\period
\end{align}
Since $K$ carries light-cone number 1, $\Ncheck$ raises the light-cone number at least 
 by 1 unit.  In this sense, $\Ncheck$ is an upper triangular matrix, while $N_q$ is diagonal. 
Due to this structure, $\Ker(N)$ is no larger than $\Ker(N_q)$, {\it i.e.} $\Ker(N_q) 
\supseteq \Ker(N)$.  To show the converse, 
 let $\ket{\psi_0}$ be any member of $\Ker(N_q)$ and construct a state  $\ket{\Psi_0}$ by 
\begin{align}
\ket{\Psi_0} &= {1\over 1+N_q\inv \Ncheck} \ket{\psi_0} 
= (1-N_q\inv \Ncheck + (N_q\inv \Ncheck)^2 \cdots)\ket{\psi_0} \period
\end{align}
$N_q\inv$ is well-defined when operated after $\Ncheck$,  since $\Ncheck$ raises the light-cone number at least by 1 unit. Operating $N=N_q+\Ncheck$ from left, one readily 
finds $N\ket{\Psi_0} 
 = N_q \ket{\psi_0} =0$, showing the inclusion $\Ker(N_q) \subseteq \Ker(N)$. 

We have now established the chain of isomorphisms:   $Q$-cohomology 
 $\simeq \Ker(N) \simeq \Ker(N_q)=\calH_T\simeq$ $Q_{-1}$-cohomology. Actually, 
 there is one more condition on   $\calH_T$. It comes from the requirement 
 $b_0\ket{\Psi} =0$. Since both $Q$ and $b_0$ annihilate the 
 physical state $\ket{\Psi}$, we must have\footnote{Even in the treatment retaining 
 the $c_0$ sector,  the same condition arises: If $L^{-,tot}_0$ does not annihilate 
a state $\ket{\Psi} \in \calH_T$, $\ket{\Psi}$ becomes $Q$-exact as 
 $\ket{\Psi}= Q\left( {1\over L^{-,tot}_0}b_0 \ket{\Psi}\right)$, which is a contradiction.}
 $L^{-,tot}_0\ket{\Psi} =0$, where  $L^{-,tot}_0\equiv  \acom{Q}{b_0}$. As $\calH_T$ 
 contains  no ghost excitations, it is equivalent to the usual ``on-shell condition" 
$L_0^-\ket{\Psi} =0$. 

Combining the result of an entirely analogous analysis for the ``$+$" sector, 
we conclude  that the physical states of the theory are 
the ones in $\calH_T$ satisfying the conditions $L^\pm_0 \ket{\Psi}=0$. 

Before concluding this subsection, we need to make a remark. In the preceding analysis, 
 we have not been  specific about the nature of the Hilbert space on which various operators 
 act. According to our phase-space normal-ordering scheme, 
the natural space would be the Fock space $\calH_{Fock}$ 
built upon the oscillator vacuum $\ket{0}$ annihilated  by the positive modes of 
$A^\star_n, B^\star_n, S^A_n$ and $B^\star_0$. 
However, as will be shown in the next subsection, non-trivial renormal-ordering will be 
required to diagonalize the on-shell conditions. This means that the physical eigenstates 
are not in $\calH_{Fock}$ and we must consider a larger Hilbert space as our arena. 
Precisely how large it should be 
 is not clear at the moment and is left for future research. Nonetheless,
 as  the cohomology analysis  itself is fairly general and its essence is simply the decoupling of
 the unphysical quartet, its validity should be independent of such uncertainty. 

\subsection{Physical spectrum and comparison with the light-cone gauge formulation}
Having  shown that the spectrum of  physical states is dictated by the $L^\pm_0$ constraints  in the transverse Hilbert space $\calH_T$ where 
 the non-zero modes of $\Pi^\pm $ and $ \Pitil^\pm$ (and all the ghosts)  are  removed, 
we now study these constraints in detail. 

Consider first the Hamiltonian constraint. Because of the redefinition (\ref{defLpmn}), 
the (dimensionless) Hamiltonian should be identified as $H=L^+_0 - L^-_0 $. In $\calH_T$ 
it  simplifies considerably and becomes quadratic in the modes. 
It takes the form (the phase-space normal-ordering is understood)
\begin{align}
H&= H_B + H_F \comma \\
H_B &= \alp \pplusup\pminusup + \half \sum (B^I_{-n} B^I_n 
+ (n^2 + M^2) A^I_{-n} A^I_{n})
\comma \\
H_F &= \half \sum (- n S^1_{-n} S^1_n +n S^2_{-n} S^2_n -iM S^1_{-n} S^2_n
+ iM S^2_{-n} S^1_n ) \period
\end{align}

Obviously the bosonic part $H_B$ describes the collection of free massive excitations 
 and one can diagonalize it with ease. For the non-zero modes, 
we introduce the following  oscillators:
\begin{align}
\altil^I_n &= {1\over \sqrt{2}} (B^I_n -i \omega_n A^I_n) \comma \qquad 
\al^I_n = {1\over \sqrt{2}} (B^I_{-n} -i \omega_n A^I_{-n}) \comma 
\end{align}
where $\omega_n$ is as defined in (\ref{defomnz}), \ie $\omega_n = (n/|n|) \sqrt{n^2+M^2}$ 
for $n\ne 0$. 
Using  the commutation relations for $A_n$ and $B_n$ oscillators, we easily verify 
\begin{align}
\com{\altil_m}{\altil_n} &= \com{\al_m}{\al_n} 
=\omega_n \delta_{m+n,0} \comma \qquad \com{\altil_m}{\al_n} =0 \period
\end{align}
Now we re-express the non-zero-mode part $H_B^{\ne 0}$ 
 in terms of these oscillators, 
and re-normal-order such that $\al_n, \altil_n$ for $n \ge 1$ are taken as 
annihiliation operators. This gives 
\begin{align}
H_B^{\ne 0} &=\sum_{n \ge 1} ( B^I_{-n} B^I_n 
+ \omega_n^2 A^I_{-n} A^I_n ) 
= \sum_{n \ge 1} (\al^I_{-n} \al^I_n+\altil^I_{-n} \altil^I_n )+8\sum_{n \ge 1} \omega_n  \comma 
\end{align}
where the last term is a divergent re-normal-ordering constant, suitably regularized. 
We shall see shortly that  this gets canceled by the fermionic contribution. 
As for the zero mode part, we identify 
\begin{align}
\al_0^I & = {1\over \sqrt{2}} (B^I_0 -iM A^I_0) \period
\end{align}
Then, $\com{\al^I_0}{{\al^J_0}^\dagger} = \delta^{IJ} M = \delta^{IJ}
\omega_0$ and the zero mode part $H_B^0$ can be rewritten as 
\begin{align}
H_B^0 &= \half ((B_0^I )^2 + M^2 (A_0^I )^2 ) = {\al_0^I }^\dagger \al_0^I 
+ 4 M \period
\end{align}
In total $H_B$ becomes 
\begin{align}
H_B &= \alp \pplusup\pminusup + {\al_0^I }^\dagger \al_0^I +
\sum_{n \ge 1} (\al^I_{-n} \al^I_n+\altil^I_{-n} \altil^I_n )
 +8\sum_{n \ge 1} \omega_n + 4M \period 
\end{align}

Next turn to  $H_F$ and again consider the non-zero mode part
 $H_F^{\ne 0}$ first. It can be written in 
 the form
\begin{align}
H_F^{\ne 0}&= \sum_{n \ge 1} ({S^1_{n}}^\dagger, {S^2_{n}}^\dagger) K(n) \vecii{S^1_n}{S^2_n} \comma 
\qquad 
K(n) = \matrixii{-n}{-iM}{iM}{n} \comma 
\end{align}
where ${S^A_n}^\dagger = S^A_{-n}$. 
For each $n$,  the hermitian matrix $K(n)$ is easily diagonalized by a unitary matrix 
$V(n)$ as 
\begin{align}
V(n)^\dagger K(n) V(n) &=  \matrixii{-\omega_n}{0}{0}{\omega_n} \period 
\end{align}
Now we define   a new basis of fermionic oscillators $S_n$ and $\Stil_n$ by 
\begin{align}
\vecii{S^\dagger_n}{\Stil_{n}} &= V^\dagger(n) \vecii{S^1_n}{S^2_n}  \period
\end{align}
The explicit expressions are
\begin{align}
S_n &= N(n) \left( S^1_{-n} -i{M \over \Omp_n} S^2_{-n} \right) 
\comma \qquad \Stil_n = N(n) \left( S^2_{n} +i{M \over \Omp_n} S^1_{n} \right) 
\comma \\
\Omp_n &\equiv \omega_n + n \comma \qquad N(n) \equiv 
 \sqrt{{\Omp_n \over \omega_n}}
\period 
\end{align}
Obviously the new oscillators satisfy the same anti-commutation relations as 
 the old ones, namely, $\acom{S_{a,m}}{S^\dagger_{b,n}} = \delta_{ab} 
\delta_{m,n}$, $\acom{\Stil_{a,m}}{\Stil^\dagger_{b,n}} = \delta_{ab} 
\delta_{m,n}$, and $\acom{S_{a,m}}{\Stil^\dagger_{b,n}} = 0$. 
Then we can rewrite $H_F^{\ne 0}$ as 
\begin{align}
H_F^{\ne 0}&= \sum_{n \ge 1} (S_n, \Stil^\dagger_n) 
\matrixii{-\omega_n}{0}{0}{\omega_n} 
\vecii{S^\dagger_n}{\Stil_{n}} =  \sum_{n \ge 1} \omega_n (-S_n S^\dagger_n
+ \Stil^\dagger_n \Stil_n) \nn\\
&= \sum_{n \ge 1}  \omega_n (S^\dagger_n S_n + \Stil_n^\dagger \Stil_n) 
 - 8 \sum_{n\ge 1} \omega_n \comma 
\end{align}
where in the last line we made a re-normal-ordering so that $S_n$ for $n \ge 1$ 
 are regarded  as annihiliation operators. Notice that the re-normal-ordering constant 
 generated by this process precisely cancels  the one produced in  the corresponding 
bosonic part, due  to supersymmetry. 

As for the zero mode part, the eigenvalues of the matrix $K(0)$ are $\pm M$ and
defining the new zero mode by 
\begin{align}
S_0 &\equiv  {1\over \sqrt{2}} (S^1_0 -iS^2_0) \comma 
\qquad \acom{S_0}{S_0^\dagger} = 1 \comma \qquad (S_0)^2 =0 \comma 
\end{align}
we can rewrite $H_F^0$ as 
\begin{align}
H_F^0 &= -iM S^1_0 S^2_0 = M S_0^\dagger S_0 - 4M \period
\end{align}
Again the re-normal-ordering constant $-4M$ cancels the corresponding contribution 
 in $H_B^0$. 

Combining all the results, we find 
\begin{align}
H &= \alp \pplusup\pminusup + {\al_0^I }^\dagger \al_0^I +
\sum_{n \ge 1} (\al^I_{-n} \al^I_n+\altil^I_{-n} \altil^I_n ) \nn\\
& + M S_0^\dagger S_0 +\sum_{n \ge 1}  \omega_n (S_n^\dagger S_n + \Stil_n^\dagger \Stil_n) \period
\end{align}
Setting this to zero and solving for $-p^-$, we precisely reproduce 
 the familiar light-cone Hamiltonian $H_{lc}$ computed in the light-cone gauge  
\cite{Metsaev:2001bj}. 

It remains to analyze  the momentum constraint, which is expressed as  $P=L^+_0 + L^-_0=0$. 
In the transverse Hilbert space $\calH_T$, $P$ reduces to 
\begin{align}
P &=  \sum_{n \ge 1}  in (A^I_{-n} B^I_n-B^I_{-n} A^I_n) +  \sum_{n\ge 1} ( n S^1_{-n} S^1_n +n S^2_{-n} S^2_n )\period
\end{align}
As was done for the Hamiltonian, we rewrite it in terms of the new oscillators. 
Again the re-normal-ordering constants cancel between the bosonic and fermionic 
 contributions and we find 
\begin{align}
P &= \sum_{n \ge 1} \left( {n \over \omega_n} \altil^I_{-n} \altil^I_n 
+ n \Stil^\dagger_n \Stil_n \right) 
-\sum_{n \ge 1} \left( {n \over \omega_n} \al^I_{-n} \al^I_n 
+ n S^\dagger_n S_n \right) \period
\end{align}
Thus, as expected,  $P=0$ yields  the  level-matching condition. 
\nullify{
Finally, let us clarify  the relation between the $sl_2$-invariant  vacuum $\ket{0}$ and 
 the light-cone type vacuum $\lcvac$ which implicitly appeared in the analysis 
 of the physical spectrum. }
\section{Discussions}
As already summarized in the introduction, we have been able to construct an exact worldsheet 
CFT description of the superstring in the plane-wave background with RR flux 
 in terms of ``free fields".  There are, however,  several issues which require further 
 understanding. 

One is the relation of our formulation to the canonical approach, described in Sec.2.2, 
which unfortunately  could not be pursued to the end due to a  technical difficulty. 
This does not of course mean that the canonical approach should be abandoned.  It would be very interesting if we can resurrect it by making use of the knowledge  of the phase-space 
formulation.  

Another point that should be clarified is the nature of the Hilbert space on which 
 the Virasoro and the BRST operators act. As remarked in Sec.4.1, this has not yet been 
fully specified. 

Let us now list some further future problems. 

The most urgent is the construction of the primary 
 operators, in particular the $(1,1)$ primaries corresponding to the low lying 
 physical excitations. Our method developed in this paper gives priority 
 to the quantization and the conformal symmetry structure and in a sense postpones 
 the real dynamical issues. As already emphasized, the dynamical properties are 
 encoded in the representation theory and by constructing the primary fields 
 we can make them explicit. In this regard, one needs to 
understand as well the basic issue of 
what are the physical quantities to be computed in this background and how they should 
 be compared to those in the super Yang-Mills theory.

Another obvious task is the understanding of the realization of the
global symmetries  of the theory\cite{Metsaev:2001bj, Metsaev:2002re}. We should construct the generators of such symmetries
 in terms of the quantized fields and check that they  close up to BRST- exact terms. 
 This would shed further  light on the justification 
 of the normal-ordering  we have adopted and the role of the quantum corrections 
 required  in the Virasoro generators.

The modular invariance issue mentioned in the introduction 
can now be addressed in the proper setting. As shown in 
\cite{Bergman:2002hv, Takayanagi:2002pi}, massive generalization of the elliptic functions 
appear and 
 the further clarification of their properties would be interesting both physically and mathematically. 

It would be an interesting project  to use our CFT description as the starting point of 
a  covariant pure spinor  formalism in operator formulation. One  way would be 
 to apply the double-spinor formalism developed in \cite{Aisaka:2005vn}, which allows 
 one to derive the pure spinor superstring  starting from a simple extension 
 of the  Green-Schwarz formalism. 

Finally, it is a  great challenge to try to apply the  phase space formalism developed here to 
some suitable  version of superstring theory in the $AdS_5\times S^5$ background. 
In principle, one should be able to quantize the theory and construct the Virasoro 
operators just as we did for the plane-wave background, since the knowledge of the 
solutions of the equations of motion is not required. Of course the analysis of the spectrum 
would be much more difficult. 

We hope to report on these and related matters in future communications. 
\par\bigskip\noindent
{\large\bf Acknowledgment}\par\smallskip\noindent
Y.K. acknowledges H.~Kunitomo for useful discussions at an  early stage of this work. 
The research of  Y.K. is supported in part by the 
 Grant-in-Aid for Scientific Research (B) 
No.~12440060 and (C) No.~18540252 from the Japan 
 Ministry of Education, Culture, Sports,  Science and Technology, while that of 
 N.Y. is supported in part by the JSPS Research Fellowships for Young Scientists.

\setcounter{equation}{0}
\renewcommand{\theequation}{A.\arabic{equation}}
\section*{Appendix A:\ \ Double-contraction contributions in 
 the Virasoro algebra with different 
normal-orderings} 
In this appendix, we display the computations of the double-contraction 
contributions in the Virasoro algebra 
 in two different normal-ordering schemes 
and clarify their differences. 

To facilitate the discussion, we first introduce  a useful  function that appears 
 in the regularization of the commutators and prepare  some formulas. 
Define the  quantity $q$ and a  function $d(q,\ep)$ by
\begin{align}
q &\equiv e^{i(\sig -\sigp)} \comma \\
d(q,\ep ) &\equiv \sum_{n \ge 0} \left(q e^{-\ep}\right)^n = {1 \over 1-qe^{-\ep}} 
\comma \\
\end{align}
where $\ep$ is an infinitesimal positive parameter. Then, it is straightforward to obtain 
 the following formulas:
\begin{align}
&\quad d(q,\ep) + d(\qinv,\ep) - 1 = 2\pi  \deltasspep \comma \\
&  \quad d(q, \ep) +d(\qinv, -\ep)  = 1\comma  \\
&  \quad d(q,\ep) -d(q,-\ep) = d(\qinv,\ep) -d(\qinv,-\ep) = 2\pi \deltasspep 
\period 
\end{align}
Here $\deltasspep$ is the usual regularized form of the $\delta$-function given by 
\begin{align}
\deltasspep &\equiv {1\over 2\pi} \sum_{n=-\infty}^\infty e^{in(\sig-\sigp)} e^{-|n|\ep} 
= {1\over \pi} {\ep \over (\sig-\sigp)^2 + \ep^2}   \period
\end{align}
Furthermore, the following formula involving the derivative of the $\delta$-function 
 will be useful:
\begin{align}
2\pi \deltasspep 
\left[d(\qinv, \ep) + d(\qinv, -\ep) \right] &=
d(\qinv, \ep)^2  -d(\qinv,-\ep)^2   \nn\\
&= 2\pi i \deltapsspep + 2\pi \deltasspep \period \label{formdeltap}
\end{align}
To derive this relation, one must expand $d(\qinv, \ep)$ up to the subleading order 
 in powers of $\sig-\sigp -i\ep$. 
\subsection*{A.1\ \  Phase-space  normal-ordering}
Consider first the commutator $\com{\calT_+(\sig)}{\calT_+(\sigp)}$ in the 
phase-space  normal-ordering. In the bosonic sector, the double-contraction contributions of interest  comes from 
\begin{align}
{1\over (2\pi)^2} \left(\com{T(\sig)}{V(\sigp)} + \com{V(\sig)}{T(\sigp)}
\right) \comma  \label{TVVT}
\end{align}
where
\begin{align}
T &= \half :\Pitil_I^2: \comma \qquad V = \chi :A_I^2: \comma \qquad 
 \chi \equiv {\muhat^2 \over 2} \Pitil^+ \Pi^+ \period 
\end{align}
Hereafter,  we will drop the transverse subscript $I$ for simplicity. The normal-ordering 
 is defined by the splitting of fields $\Pitil=\Pitil\toplus + \Pitil\tominus$ and $A
=A\toplus + A\tominus$, where the superscript $(+)$ ( $(-)$) denotes
 the annihiliation (creation) part.  In the phase-space
 normal-ordering, we have
\begin{align}
\Pitil\toplus(\sig) &= \sum_{n\ge 0} \Pitil_n e^{-in\sig} 
\comma \qquad \Pitil\tominus (\sig) = \sum_{n \ge 1} \Pitil_{-n} e^{in\sig}  
\comma 
\label{Pitilsplit}\\
A\toplus(\sig) &= \sum_{n \ge 1} A_n e^{-in\sig} \comma \qquad 
 A\tominus(\sig) = \sum_{n \ge 0} A_{-n} e^{in\sig} \period
\end{align}
The commutator $\com{T(\sig)}{V(\sigp)} $ is defined by 
$T(\sig-i\ep)V(\sigp) -V(\sigp-i\ep) T(\sig)$. It is easy to show that this amounts to 
taking the usual commutator once and then normal-order the remaining  operator product. 
We thus obtain
\begin{align}
\com{T(\sig)}{V(\sigp)} &= -{2\pi i \over \sqrt{2}} \chi(\sigp) \deltasspep
(\Pitil(\sig-i\ep) A(\sigp) + A(\sigp-i\ep)\Pitil(\sig))  \nn\\
&= -{\pi i \over \sqrt{2}} \chi(\sigp) \deltasspep :\Pitil(\sig)A(\sig):  \nn\\
&-{2\pi i \over \sqrt{2}} \chi(\sigp) \deltasspep \left(\com{\Pitil\toplus(\sig-i\ep)}{A\tominus(\sigp)} + \com{A\toplus(\sigp-i\ep)}{\Pitil\tominus(\sig)} \right) 
\period \nn
\end{align}
The commutators in the second line are given by
\begin{align}
\com{\Pitil\toplus(\sig-i\ep)}{A\tominus(\sigp)}  &= -{i \over \sqrt{2}} d(\qinv,\ep) \comma 
\label{PitilpAm}\\
\com{A\toplus(\sigp-i\ep)}{\Pitil\tominus(\sig)} &= -{i \over \sqrt{2}} d(\qinv,-\ep) 
\period \label{ApPitilm} 
\end{align}
We are interested in the double-contraction (DC)  part given in the second line. 
Using the formula (\ref{formdeltap}),  we readily  obtain
\begin{align}
\com{T(\sig)}{V(\sigp)}_{DC} &= -\pi i \chi(\sigp)  \deltapssp  - \pi \chi(\sig) \deltassp
\period
\label{TV} 
\end{align}
$\com{T(\sigp)}{V(\sig)} $ can be  obtained  from this by the interchange $\sig 
\leftrightarrow \sigp$.  

Combining these results,  the operator parts cancel and we get 
\begin{align}
\com{T(\sig)}{V(\sigp)}
+\com{V(\sig)}{T(\sigp)}= -\pi i  \left[ 2\chi(\sig) \deltapssp 
+ \del_\sig \chi(\sig) \deltassp \right]  \period \label{dcboson}
\end{align}

Now consider the contribution in the fermionic sector. The double-contraction contribution
 of interest comes from 
\begin{align}
{1\over (2\pi)^2} \com{F(\sig)}{F(\sigp)} \comma 
\end{align}
where
\begin{align}
F(\sig) &= \rho S^1(\sig) S^2(\sig) \comma \qquad \rho \equiv -{i \muhat \over \sqrt{2}}
\sqrt{\Pitil^+ \Pi^+} \period
\end{align}
Note the important relation $\rho^2 = -\chi$.  The normal-ordering of the fermion $S^A$
 is defined by the splitting
\begin{align}
S^A (\sig) &= {S^A}\toplus(\sig) + S^A_0 + {S^A}\tominus (\sig) \comma \\
{S^A}\toplus(\sig) &= \sum_{n \ge 1} S^A_n e^{-in\sig} 
\comma \qquad {S^A}\tominus (\sig) = \sum_{n \ge 1} S^A_{-n} e^{in\sig} 
\period
\end{align}
Defining the regularized commutator in the same way as in the case of the bosons, 
we get
\begin{align}
\com{F(\sig)}{F(\sigp)} & = 2\pi \rho(\sig) \rho(\sigp) 
\deltasspep \left( -S^2(\sig-i\ep) S^2(\sigp) 
 + S^1(\sigp-i\ep) S^1(\sig)  \right)\period 
\end{align}
The products appearing here are normal-ordered as 
\begin{align}
S^2(\sig-i\ep) S^2(\sigp)  &= 
:S^2(\sig) S^2(\sigp): + \,   \half + \acom{{S^2}\toplus(\sig-i\ep)}{{S^2}\tominus(\sigp)}\comma \\
 S^1(\sigp-i\ep) S^1(\sig)  &=
:S^1(\sigp) S^1(\sig): +\,    \half + \acom{{S^1}\toplus(\sigp-i\ep)}{{S^1}\tominus(\sig)} 
\comma 
\end{align}
where the term $1/2$ is from the zero-mode and the anti-commutators are given by 
\begin{align}
\acom{{S^2}\toplus(\sig-i\ep)}{{S^2}\tominus(\sigp)} &= d(\qinv, \ep)-1  \comma 
\label{Stwopm}\\
\acom{{S^1}\toplus(\sigp-i\ep)}{{S^1}\tominus(\sig)} &= d(q, \ep)-1  \period
\label{Sonepm}
\end{align}
The  normal-ordered products  such as $:S^2(\sig)S^2(\sigp):$ vanish at $\sig=\sigp$
and hence do not contribute in  the presence of $\deltasspep$ as above. 
Using the formulas for the $d(q,\ep)$ function and the relation $\rho^2=-\chi$, 
we then get
\begin{align}
\com{F(\sig)}{F(\sigp)} &= 2\pi \rho(\sig)^2 \deltasspep 
 -2\pi \rho(\sig)\rho(\sigp) \deltasspep (d(\qinv, \ep)+ d(\qinv, -\ep)) \nn\\
&= \pi i \left( 2\chi(\sig) \deltapssp + \del_\sig\chi(\sig) \deltassp \right)
\period
\end{align}
This precisely cancels the contribution (\ref{dcboson}) from the bosonic sector. 

Computations for $\com{\calT_-(\sig)}{\calT_-(\sigp)}$ and
 $\com{\calT_+(\sig)}{\calT_-(\sigp)}$  are quite similar and the double-contractions 
 again cancel between the bosonic and fermionic contributions. Hence, {\it the Virasoro 
 algebra properly closes in the phase-space normal-ordering. }
\subsection*{A.2\ \  Massless  normal-ordering}
Again we  begin with the bosonic sector of the 
commutator $\com{\calT_+(\sig)}{\calT_+(\sigp)}$ and focus on the expression 
 (\ref{TVVT}). The difference from the previous phase-space normal-ordering is that the 
field $A$ must be split into the  annihiliation and the creation parts in the 
 following way:
\begin{align}
A(\sig) &= {\calA}\toplus(\sig) + {\calA}\tominus(\sig) \comma \\
{\calA}\toplus(\sig) &=    {i \over \sqtwo} \sum_{n\ge 1} {1\over n}
 \left( \Pitil_n e^{-in\sig} + \Pihat_n e^{in\sig} \right) \comma \\
{\calA}\tominus(\sig)&= A_0-{i \over \sqtwo} \sum_{n\ge 1} {1\over n}
 \left( \Pitil_{-n} e^{in\sig} + \Pihat_{-n} e^{-in\sig} \right) \period
\end{align}
The split for $\Pitil$ is as before, namely (\ref{Pitilsplit}). Denoting this  normal-ordering
 by $\ant \quad \ant$, $\com{T(\sig)}{V(\sigp)}$ becomes
\begin{align}
\com{T(\sig)}{V(\sigp)}
&= -{\pi i \over \sqrt{2}} \chi(\sigp) \deltasspep \ant\Pitil(\sig)A(\sig)\ant  \nn\\
&-{2\pi i \over \sqrt{2}} \chi(\sigp) \deltasspep \left(\com{\Pitil\toplus(\sig-i\ep)}{{\calA}\tominus(\sigp)} + \com{\calA\toplus(\sigp-i\ep)}{\Pitil\tominus(\sig)} \right) 
\period \nn
\end{align}
The commutators in the last line are given by 
\begin{align}
\com{\Pitil\toplus(\sig-i\ep)}{\calA\tominus(\sigp)}  &= -{i \over \sqrt{2}} d(\qinv,\ep) 
\comma 
\nn\\
\com{\calA\toplus(\sigp-i\ep)}{\Pitil\tominus(\sig)} &= -{i \over \sqrt{2}} d(\qinv,-\ep) 
\comma \nn
\end{align}
which are identical to (\ref{PitilpAm}) and (\ref{ApPitilm}).  Therefore the rest of the 
 calculations are also as before and we obtain 
\begin{align}
\com{T(\sig)}{V(\sigp)}
+\com{V(\sig)}{T(\sigp)}= -\pi i  \left[ 2\chi(\sig) \deltapssp 
+ \del_\sig \chi(\sig) \deltassp \right]  \period
\end{align}
This  coincides with (\ref{dcboson}). 

We now  turn to the fermionic sector.  Although the operator $F(\sig)$ itself does not require 
 normal-ordering, we must  interpret $F(\sig)$ as $F(\sig) = \rho(\sig) \Shat^1(-\sig) S^2(\sig)$, where $\Shat^1(-\sig)$ is split into 
\begin{align}
\Shat^1(-\sig) &= \Shatonep(-\sig) + S^1_0 
+ \Shatonem(-\sig) \comma \\
\Shatonep(-\sig) &= \sum_{n \ge 1} \Shat^1_n e^{in\sig} 
\comma \qquad  \Shatonem(-\sig) = \sum_{n\ge 1} \Shat^1_{-n} e^{-in\sig}
\period 
\end{align}
The splitting  for $S^2$ is the same as in the phase-space normal-ordering.
Then the commutator $\com{F(\sig)}{F(\sigp)}$ becomes 
\begin{align}
\com{F(\sig)}{F(\sigp)}&= 2\pi \rho(\sig) \rho(\sigp) 
\deltasspep \left( -S^2(\sig-i\ep) S^2(\sigp) 
 + \Shat^1(-\sigp-i\ep) \Shat^1(-\sig)  \right) \period \label{FF} 
\end{align}
Now $\Shat^1(-\sigp-i\ep) \Shat^1(-\sig) $ must be normal-ordered as
\begin{align}
\Shat^1(-\sigp-i\ep) \Shat^1(-\sig) &= \ant \Shat^1(-\sigp) \Shat^1(-\sig) \ant
+ \half + \acom{\Shatonep(-\sigp -i\ep)}{\Shatonem(-\sig)} \comma \nn
\end{align}
where the anticommutator is given by 
\begin{align}
\acom{\Shatonep(-\sigp -i\ep)}{\Shatonem(-\sig)} &= d(\qinv, \ep) -1 \period
\end{align}
The crucial difference from the phase-space
 normal-ordering is that,  contrary to (\ref{Sonepm}), 
 $\acom{\Shatonep(-\sigp -i\ep)}{\Shatonem(-\sig)} $ is identical to 
$\acom{{S^2}\toplus(\sig-i\ep)}{{S^2}\tominus(\sigp)}$ given in (\ref{Stwopm}). 
Therefore the contributions from $\Shat^1$ and $S^2$ cancel in (\ref{FF}) and 
we get $\com{F(\sig)}{F(\sigp)} =0$. 

Thus, in the massless normal-ordering scheme, the double-contraction contributions 
 from the bosons and the fermions {\it do not} cancel and $\com{\calT_+(\sig)}{\calT_+(\sigp)}$
contains an extra operator anomaly besides the usual c-number anomaly. Reinstating the 
 factor of $8$ (due to  the transverse degrees of freedom), it reads
\begin{align}
\com{\calT_+(\sig)}{\calT_+(\sigp)}_{extra} &= 
 -{2 i \over \pi} (2\chi(\sig) \deltapssp +\del_\sig\chi(\sig)\deltassp) \period
\end{align}

The computation of $\com{\hat{\calT}_-(\sig)}{\hat{\calT}_-(\sigp)}$ is similar 
 and the result again contains the operator anomaly:
\begin{align}
\com{\hat{\calT}_-(\sig)}{\hat{\calT}_-(\sigp)}_{extra} &= 
-{2 i \over \pi} (2\hat{\chi}(\sig) \deltapssp +\del_\sig\hat{\chi}(\sig)\deltassp) 
\comma 
\end{align}
where $\hat{\chi}(\sig) = \chi(-\sig) = \Pitilhat^+\Pihat^+$. 

Finally, consider the commutator $\com{\calT_+(\sig)}{\hat{\calT}_-(\sigp)}$. 
Actually, to make use of the various formulas already developed, it is more convenient
 to compute $\com{\calT_+(\sig)}{\hat{\calT}_-(-\sigp)}$. 
For this quantity, while the fermionic contribution for the double-contraction part 
 continues to vanish, a new situation occurs for the bosonic contribution. The relevant 
part is 
\begin{align}
{1\over (2\pi)^2} \left( \com{T(\sig)}{\hat{V}(-\sigp)} 
- \com{\hat{T}(-\sigp)}{V(\sig)}\right) \comma 
\end{align}
The double-contraction part of $\com{T(\sig)}{\hat{V}(-\sigp)}$ is easily seen to be 
 the same as  given in (\ref{TV}), namely
\begin{align}
\com{T(\sig)}{\hat{V}(-\sigp)}_{DC} &= -\pi i \chi(\sigp)  \deltapssp  - \pi \chi(\sig) \deltassp
\period 
\end{align}
On the other hand, $\com{\hat{T}(-\sigp)}{V(\sig)}$
is not obtained simply by making the interchange  $\sig \leftrightarrow \sigp$. 
Using the additional formulas
\begin{align}
\com{{\Pihat}\toplus(-\sigp)}{{\calA}\tominus(\sig)} &= -{i \over \sqrt{2}}
d(\qinv, \ep) \comma \nn\\
\com{{\calA}\toplus(\sig)}{{\Pihat}\tominus(-\sigp)} &= -{i \over \sqrt{2}}
d(\qinv, -\ep) \comma  \nn
\end{align}
we get
\begin{align}
\com{\hat{T}(-\sigp)}{V(\sig)}_{DC}
&=  -\pi i \chi(\sig)  \deltapssp  - \pi \chi(\sig) \deltassp \period
\end{align}
Note that it is identical to $\com{T(\sig)}{\hat{V}(-\sigp)}$ above, except 
for the argument of $\chi$ in front of $\deltapssp$. Due to this difference, they do not quite cancel and  produce an  operator anomaly. Flipping the sign of $\sigp$
 and supplying numerical factors, it is given by
\begin{align}
\com{\calT_+(\sig)}{\hat{\calT}_-(\sigp)}
&= -{2 i \over \pi} \del_\sig \chi(\sig) \delta (\sig + \sigp) \period
\end{align}

\end{document}